\documentclass[12pt]{article}
\usepackage{epsfig}
\usepackage{amsbsy}
\usepackage{amsfonts}

\def\bea{\begin{eqnarray}}
\def\eea{\end{eqnarray}}
\def\beqn{\begin{eqnarray}}
\def\eeqn{\end{eqnarray}}
\def\beq{\begin{equation}}
\def\eeq{\end{equation}}

\def\l{\left}
\def\r{\right}
\def\ie{{\it i.e.}}
\def\eg{{\it e.g.~}}

\def\Dslash{\not{\hbox{\kern-4pt $D$}}}
\def\pslash{\not{\hbox{\kern-4pt $p$}}}

\def\lan{\langle}
\def\ran{\rangle}
\def\blan{\Big\langle}
\def\bran{\Big\rangle}

\def\lan{\langle}
\def\ran{\rangle}

\def\pr{{\it Phys.~Rev.~}}

\def\np{{\it Nucl.~Phys.~}}

\def\pl{{\it Phys.~Lett.~}}

\def\vol#1{{\bf #1}}
\def\vyp#1#2#3{\vol{#1} (#2) #3}

\def\ngen{rep}

\def\be{\begin{equation}}
\def\ee{\end{equation}}
\def\bea{\begin{eqnarray}}
\def\eea{\end{eqnarray}}
\def\beq{\begin{equation}}
\def\eeq{\end{equation}}
\def\bq{\begin{quote}}
\def\eq{\end{quote}}
\def\gappeq{\mathrel{\rlap {\raise.5ex\hbox{$>$}} {\lower.5ex\hbox{$\sim$}}}}
\def\lappeq{\mathrel{\rlap{\raise.5ex\hbox{$<$}} {\lower.5ex\hbox{$\sim$}}}}

\def\vol#1{{\bf #1}}
\def\vyp#1#2#3{\vol{#1} (#2) #3}

\def\epm#1#2{\hbox{${\lower1pt\hbox{$\scriptstyle +#1$}}
\atop {\raise1pt\hbox{$\scriptstyle -#2$}}$}}

\def\gsim{\mathrel{\rlap{\lower4pt\hbox{\hskip1pt$\sim$}}
    \raise1pt\hbox{$>$}}}         
\def\eg{{\it e.g.}}

\def\frac#1#2{{{#1}\over {#2}}}

\def\bq{\bar{q}}
\catcode`@=11 
\def\slash#1{\mathord{\mathpalette\c@ncel#1}}
 \def\c@ncel#1#2{\ooalign{$\hfil#1\mkern1mu/\hfil$\crcr$#1#2$}}
\def\lsim{\mathrel{\mathpalette\@versim<}}
\def\gsim{\mathrel{\mathpalette\@versim>}}
 \def\@versim#1#2{\lower0.2ex\vbox{\baselineskip\z@skip\lineskip\z@skip
       \lineskiplimit\z@\ialign{$\m@th#1\hfil##$\crcr#2\crcr\sim\crcr}}}
\catcode`@=12 

\begin{document} 
\pagestyle{empty} 
\begin{flushright} 
{\tt hep-ph/0204232}\\{GeF/TH/3-02}\\
{RM3-TH/02-01}\end{flushright} \vspace*{5mm}
\begin{center} {\Large\bf Neural Network Parametrization of
Deep--Inelastic Structure Functions}\\
\vspace*{0.8cm} {\bf 
{\bf Stefano Forte$^{a\;}$}, 
Llu\'\i s Garrido$^{b\;}$, Jos\'e I. Latorre$^{b\;}$ and Andrea Piccione$^{c\;}$}
 \\\vspace{0.6cm}
{}$^a$INFN, Sezione di Roma Tre\\
Via della Vasca Navale 84, I-00146 Rome, Italy\\
\vspace{0.4cm}
{}$^b$Departament d'Estructura i Constituents de la Mat\`eria,
Universitat de Barcelona,\\
Diagonal 647, E-08028 Barcelona, Spain\\
\vspace{0.4cm}
{}$^c$INFN sezione di Genova and Dipartimento di Fisica, 
Universit\`a di Genova,\\
via Dodecaneso 33, I-16146 Genova, Italy\\

\vspace*{0.9cm}
{\bf Abstract}
\end{center}
\noindent
We construct  a parametrization of 
deep--inelastic 
structure functions  which retains information on
experimental errors and correlations, and which does not  introduce
any theoretical bias while interpolating between existing data points.
We generate a Monte Carlo sample of pseudo--data configurations and we
train an ensemble of neural networks on them. This effectively
provides us with a probability measure in the space of structure
functions, within the whole kinematic region where data are
available. This measure can then be used to determine the value of the
structure function, its error,  point--to--point correlations and
generally the value and uncertainty of any function of the structure
function itself. We
apply this technique to the determination of the structure function
$F_2$ of the proton and deuteron, and a precision determination of the
isotriplet combination $F_2$[p-d]. We discuss in detail these results, 
check their stability and accuracy, and make them
available in various formats for applications.  \\
\vspace*{0.5cm}

\vfill
\noindent

\begin{flushleft}April 2002 \end{flushleft} 
\eject

\setcounter{page}{1} \pagestyle{plain}
\section{Determining Structure Functions}
Recently, a considerable amount of theoretical and experimental effort
has been invested in the accurate determination of the parton
distributions of the nucleon, and in particular their associate
errors, in view of the accurate computation of collider processes
and determination of QCD parameters~\cite{qcdrev}--\cite{cteq6}. 
The determination of parton distributions requires
unfolding the experimentally measurable structure functions in terms
of their parton content, by resolution of the QCD evolution equations.
However, for many applications, it is desirable and sufficient to deal
directly with the experimentally measurable structure functions
themselves. For instance, the nonsinglet structure function $F_2$
coincides with the (DIS--scheme) quark distribution, and thus in
particular the strong coupling can be extracted directly from its
scaling violations~\cite{trmom}.  
Also, knowledge of the structure function $F_2$
is needed for data analysis: for instance, when extracting polarized
structure functions from cross--section asymmetries~\cite{smc}.

For all these applications, one needs to know the structure function
$F_2$ as a function of $x$ and $Q^2$, and not only for the particular
values where it has been measured. Therefore, even though the
laborious task of disentangling the parton content of the target and
solving the evolution equations is not necessary if one directly 
determines the
structure function, the central problem is the same as in the determination
of parton
distributions. Namely, one has to determine a function, and associate error,
from a finite set of measurements. Because a function is an
infinite--dimensional object, this is an a priori ill--posed problem, unless
one makes some extra assumptions. The simplest way of implementing
suitable extra assumption is to assume 
that the given function has a certain functional form, parametrized by a
finite number of parameters, which can then be fitted from the
data. However, the choice of a specific functional form is in general
quite restrictive: therefore, it
is thus a source of bias, \ie\ systematic error which is very difficult to
control. Furthermore, when fitting a fixed functional form to the
data, it is very hard to obtain a determination not only of the
best-fit parameters, but also of their errors (and correlations):
systematic uncertainties are difficult to take into
account~\cite{alekhin,cteq6}, and
conventional error propagation and determination of the covariance
matrix in a fixed parameter space is unwieldy and may fail because of
linearization problems. 

A way out of these difficulties, based on the direct determination of
the measure in the space of parton distributions by Monte Carlo
sampling has been proposed~\cite{GKK,GKK2}. 
However, it is computationally quite
intensive, and it has not yet led to a full set of parton
distributions with errors.
Here, we propose a new approach to the determination of structure
functions which also aims at the determination of the measure in the
space of structure functions. Our approach is based on the use of
neural networks as basic interpolating tools. The peculiar feature of
neural networks which we will exploit is that they can provide an
interpolation without requiring any assumption (other than continuity)
on the functional form of the underlying law fulfilled by the given
observable. Hence, given a sampling of the measure in the space of
function over a finite set of points, the neural networks can provide
an unbiased interpolation which gives us the measure for all points,
at least within a range of $x$ and $Q^2$ where the sampling provided
by the data is fine enough.
 
We will use this procedure to construct parametrizations of the
structure function $F_2$ for the proton, the deuteron, and an
independent precision determination of the nonsinglet combination
$F_2^p-F_2^d$. We will analyze in detail the properties of our
results, and show that they reproduce correctly the information
contained in the data, while interpolating smoothly between data
points.  We will also see that when the
separation of data points in $x$ and $Q^2$ is smaller than a certain
correlation length, the neural net manages to combine the
corresponding experimental information, thus leading to a
determination of the structure function which is more accurate than
that of individual experiments, and we will devise statistical tests
to ascertain that this is achieved in an unbiased way.

In Sect.~2 we will introduce the problem of constructing a
parametrization of structure functions, the data that we will use,
discuss previous solutions of this problem, and motivate our
alternative approach.
In Sect.~3 we will then provide some general background on
neural networks, and in Sect.~4 we will describe our method for the
construction of a neural parametrization of structure functions.
Finally, in Sect.~5 we will turn to a discussion and assessment
of our results. First
we will discuss the theoretically simpler nonsinglet structure
function: we will examine the compatibility of different experiments,
and the ability of the neural networks to combine different pieces of
experimental information. Then, we will tackle the determination of
the individual proton and deuteron structure functions, and discuss
the accuracy of our results.

\section{Parametrization of Structure functions}

Our aim in this paper is to construct a simultaneous parametrization
of the structure functions $F_2$ for the proton and deuteron: we start
from a set of measurements of a structure function, say $F_2$, for
a discrete set of values of $x$ and $Q^2$, and we wish to determine
$F_2 (x, Q^2)$ as a function defined in a certain range of $x$ and
$Q^2$.  We choose the structure function $F_2$ as determined in
neutral--current DIS of charged leptons, since an especially wide set of
experimental determinations of it is available, and also because it is
of direct theoretical interest. Studies of charged--current, neutrino,
and polarized structure functions will be left to future work.

A simultaneous determination of the structure functions for proton and
deuteron is very useful for applications, since it amounts to a
simultaneous determination of the isospin singlet and triplet
components of the structure function.  Also, such a simultaneous
determination poses peculiar theoretical problems, given that we aim
at a determination of the full set of correlations.

In fact, we will provide both a determination of the correlated pair
of proton and deuteron structure functions, as well as a separate
determination of the isotriplet component. This is both useful and
interesting because, as we shall see, data on the isotriplet
combination turn out to be affected by much larger percentage errors,
\ie~to have a much less favourable signal--to--noise ratio: the
isotriplet is generally a small number obtained as the difference of
two numbers which are by an order of magnitude larger. Therefore, on
the one hand a more precise determination can be obtained by a
dedicated analysis, rather than just taking a difference, on the other
hand this determination will turn out to have peculiar features which
deserve a specific study.

Parametrizations of structure functions have been constructed before,
and used for the applications discussed in the introduction. In this
section, we will first describe the specific data set that we will use,
and then briefly review existing approaches to the parametrization of
structure functions and their shortcomings.

\subsection{Experimental data}

\begin{figure}[t]
\begin{center}
\epsfig{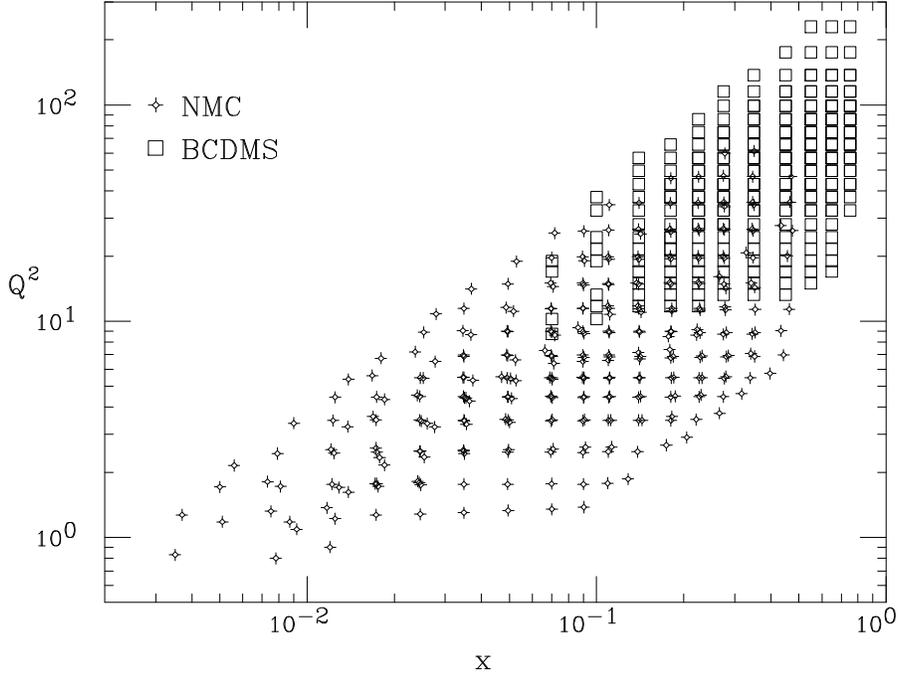}
\end{center}
\begin{center}
\caption{NMC and BCDMS kinematic range.}
\label{fig:kinrange}
\end{center}
\end{figure}

\noindent
We will use the data of the New Muon Collaboration (NMC)~\cite{NMC}
and the BCDMS (Bologna-CERN-Dubna-Munich-Saclay)
Collaboration~\cite{BCDMS} on the structure function $F_2$ of proton
and deuteron. These data provide a simultaneous determination of the
proton and deuteron structure functions in the same kinematic region,
and provide the full set of correlated experimental systematics for
these measurements. Earlier data from SLAC are not competitive with
these in terms of accuracy and kinematic coverage. 
 The more recent
HERA data are available in a much wider kinematic region, but only for
proton targets. Another set of joint proton and deuteron measurements
was performed by the E665 Collaboration \cite{e665}. These data are
mostly concentrated at low $Q^2$ (and low $x$): 80\% of the data have
$Q^2< 1$~GeV$^2$. They are therefore somewhat less interesting for
perturbative QCD applications, and we will leave them out of the
present analysis. We have checked that their inclusion would have a
negligible impact on the determination of the structure function in
the region covered by NMC and BCDMS.  The kinematic coverage of the
data which we include in our analysis is displayed in
Figure~\ref{fig:kinrange}. A better coverage of the large $x$ region
at moderate $Q^2$ might be achievable in the near future by
combining some of the earlier SLAC data~\cite{slac} with forthcoming JLAB
data~\cite{jlab}. 

\subsubsection{NMC}

The NMC data consist of four data sets for the proton and the deuteron
structure functions corresponding to beam energies of $90$~GeV ($72$ p
and $72$ d data points), 120~GeV ($64+64$ data points), $200$~GeV
($74+74$ data points), and $280$~GeV ($78+78$ data points). They cover
the kinematic range $0.002\le x\le 0.60$ and $0.5\,\rm{GeV}^2\le Q^2
\le 75\,\rm{GeV}^2$.  The systematic errors, given in
Ref.~~\cite{NMC}, are:
\begin{itemize}
\item uncertainty on the incoming and outgoing beam energies, fully
correlated between proton and deuteron data, but independent for data
taken at different beam energies ($E$, $E'$);
\item radiative corrections, fully correlated between all energies,
but independent for proton and deuteron ($RC$);
\item acceptance ($AC$) and reconstruction efficiency ($RE$) , fully
correlated for all data sets;
\item normalization uncertainty, correlated between the proton and the
deuteron data, but independent for data taken at different beam
energies ($\sigma_N$).
\end{itemize}
In this experiment, correlation between proton and deuteron data are
due to the fact that both targets were exposed to the beam simultaneously.

The uncertainties due to acceptance range from 0.1 to 2.5\% and reach
at most 5\% at large $x$ and $Q^2$. The uncertainty due to radiative
corrections is highest at small $x$ and large $Q^2$ and is at most
2\%.  The uncertainty due to reconstruction efficiency is estimated to
be 4\% at most. The uncertainties due to the incoming and the
scattered muon energies contribute to the systematic error by at most
2.5\%. The normalization uncertainty is 2\%.

\subsubsection{BCDMS}

The BCDMS data consist of four data sets for the proton structure
function, corresponding to beam energies of $100$~GeV (97 data
points), $120$~GeV (99 data points), $200$~GeV (79 data points) and
$280$~GeV (76 data points), and three data sets for the deuteron
structure function corresponding to beam energies of $120$~GeV (99
data points), $200$~GeV (79 data points) and $280$~GeV (76 data
points).  They cover the kinematic range $0.06\le x\le 0.80$ and $7
\,\rm{GeV}^2\le Q^2 \le 280\,\rm{GeV}^2$.

The systematic errors are:~\footnote[1]{Note that the full set of
systematics is listed in the preprint version 
but not in the published version of Ref.~\cite{BCDMS}.}
\begin{itemize}
\item calibration of the incoming muon (beam) energy ($f_b$);
\item calibration of the outgoing muon energy (spectrometer 
magnetic field) ($f_s$);
\item spectrometer resolution ($f_r$);
\item absolute normalization uncertainty ($\sigma_{N_a}$);
\item relative normalization uncertainties ($\sigma_{N_t}$,
$\sigma_{N_b}$).
\end{itemize}
All these sources of systematics are fully correlated for all targets
and for all beam energies, despite the fact that measurements with
different targets were done at different times, for
the following reasons~\cite{milsztajn}:
the calibration of the incoming beam energy E was dominated by a
systematic uncertainty which was more stable in time than the precision of
measurements;
the calibration of the outgoing muon energy was reproducible
to high relative accuracy throughout the experiment, independent of the
target or beam energy;
the resolution of the spectrometer depended on the constituent
material, which did not change during the running of the experiment.

The uncertainty due to beam energy is smaller than 5~\%, and those due
to magnetic field and outgoing muon energy are smaller than 10~\%; all
uncertainties are most relevant at large $x$ and small $Q^2$.  The
absolute cross--section normalization error is 3\%; relative
normalization errors are $2\%$ between data taken with different
targets, $1\%$ between data taken at different beam energies for the
proton, $1\%$ between data taken at 120 GeV and 200 GeV and $1.5\%$
between data taken at 200 GeV and 280 GeV for the deuteron.

\subsubsection{Correlation and covariance matrices}

The structure of the correlation matrix, expressed in
 terms of the sources of systematics discussed above is as follows.
 For NMC 
\bea \label{cornmc}
\rho_{ij} &=&\frac{F_i F_j }
 {\sigma_{i,tot} \sigma_{j,tot} } \nonumber \\
 &\times&\left( E_i\,E_j+E'_i\,E'_j+AC_i\,AC_j+ RC_i\,RC_j+RE_i\,RE_j
 +\sigma_{N}^2\right) 
\eea 
where
 $\sigma_{tot} =\sqrt{
 {\sigma _s}^2+F 
(\sigma_c^2+\sigma_N^2)}$; $\sigma_s $ is the
 statistical error, $\sigma_c $ is the combination of
 correlated systematic errors, 
$\sigma_N$ is the normalization error, and  systematic and
 normalization errors are given in percentage.
 
For BCDMS 
\beq 
\rho_{ij} =\frac{F_i F_j }
{\sigma_{i,tot} \sigma_{j,tot} }
\left(f_{b,i}\,f_{b,j}+f_{s,i}\,f_{s,j}+f_{r,i}\,f_{r,j}
+\sigma_{N}^2\right) 
\label{corbcd}
\eeq with
$\sigma_{N}=\sqrt{\sigma_{N_a}^2+\sigma_{N_t}^2+\sigma_{N_b}^2}$,
where $\sigma_{N_a}$ is a global normalization error, 
$\sigma_{N_t}$ is the relative normalization between different
targets and $\sigma_{N_b}$ is the relative normalization between
different beam energies.

The covariance matrix is defined as
\beq
\mathrm{cov}_{ij}=\rho_{ij}\sigma_i\sigma_j .
\label{covdef}
\eeq
In the sequel, we will also be interested in the covariance matrix
of data for the nonsinglet structure function $F^{NS}\equiv F^p-F^d$.
In general the covariance of two functions of $x_i$, $x_j$ is related
to the covariance of these variables by
\beq
\mathrm{cov}(f(\vec x),g(\vec x))=\sum_{i,j}
\frac{\partial f}{\partial x_i}\frac{\partial g}{\partial x_j}
\mathrm{cov}(x_i,x_j)\ .
\label{covfun}
\eeq
For the nonsinglet structure function we get
\beq
\mathrm{cov}(F^p_i-F^d_i,F^p_j-F^d_j)=
\mathrm{cov}(F^p_i,F^p_j)+\mathrm{cov}(F^d_i,F^d_j)-
\mathrm{cov}(F^p_i,F^d_j)-\mathrm{cov}(F^d_i,F^p_j)\ .
\label{covdif}
\eeq

\subsection{Structure function fits}

\noindent

Several structure function fits have been presented in the
literature
~\cite{smc,e665,tulay}. They are all based on the idea of
assuming a more or less simple functional from for the structure
function $F_2$ (say), and then fitting its parameters.  Once the free
parameters of a given functional form are determined by fitting to the
data, one can compute any function of $F_2$, such as \eg\ its Mellin
moments. In principle, given the error matrix of the parameters of the
fit, one can also determine errors on $F_2$ and functions thereof by
error propagation.  Clearly, however, this is both impractical and
subject to errors which are difficult to assess, because of the
numerous linearizations which may be needed in order to determine the
error on a physical observable (such as a Mellin moment) and, more
importantly, because of the bias which is imposed by the choice of a
specific functional form.
 
In order to overcome at least the practical difficulty of determining
errors by propagation from the covariance matrix, a frequently adopted
shortcut consists of giving a band of acceptable results, 
\ie~a pair
of `upper' and `lower' values of the structure function, for each
$(x,\,Q^2)$. This error band can be again calculated either by means
of the covariance matrix itself~\cite{smc,tulay}, or by some other recipe
which involves assessing the compatibility of a certain result with
the data~\cite{cteq6}. Whereas this may be sufficient in order to
estimate the size of the error related to uncertainties on structure
functions, it is clearly inadequate for a quantitative analysis, since
there is no way to use these error bands to uniquely determine the
error on a given quantity which depends on the structure function. For
example, suppose one wants to determine the error on the integral over $x$
of the structure function: one might be tempted to identify this error
with the spread of values of the integral computed from the upper and
lower curves. This is however not correct, because it neglects the
correlation between the structure function at different values of $x$,
and it can in fact lead to a substantial overestimate of the error.

These difficulties can be traced to the fact that when fitting a
structure function one is trying to determine a function, \ie~an
infinite--dimensional quantity, from a finite set of
measurements~\cite{GKK}. Assuming a functional form of the function
projects the problem onto the finite--dimensional subspace of
functions of the given form, but at the cost of introducing a bias
when choosing how to perform this projection. A full, unbiased
solution of the problem requires instead the determination of the
probability measure ${\cal P}[F_2]$ in the space of functions
$F_2(x,Q^2)$, such that the expectation value of any observable ${\cal
F} \left[F_2(x,Q^2)\right]$ which
depends on $F_2$ can be found by
averaging with this measure:
\begin{equation}
\blan {\cal F} \left[F_2(x,Q^2)\right]\bran=\int\! {\cal D} F_2\, {\cal F} \left[F_2(x,Q^2)\right]\, {\cal P}[F_2],
\label{funave}
\end{equation}
where the integral is a functional integral over the space of
functions $F_2$~\cite{GKK}. For example, ${\cal
F} \left[F_2(x,Q^2)\right]$ could be a Mellin moment of $F_2(x,Q^2)$.

A practical way of determining this probability measure, in the context
of determining parton distributions, has been suggested in
Ref.~\cite{GKK2}. The measure is determined by means of a Monte Carlo
approach coupled to Bayesian inference: one builds a Monte Carlo
sample of the space of functions, starting from an essentially
arbitrary assumption on the form of the measure, and then one applies
Bayesian inference in order to update this measure based on the
available data.

Here, we suggest a different strategy to attain the same goal, in the
related context of fitting structure functions: we use the available
data to generate a Monte Carlo sample of the functional measure at a
discrete set of points, and then use neural networks to interpolate
between these points, thus generating the desired representation
of the probability measure in the space of functions.  After
introducing in the next section the general features of neural
networks, we will give a detailed explanation of our procedure in
Sect.~4.

\section{Neural networks}

Artificial neural networks provide unbiased robust universal
approximants to incomplete or noisy data.  Applications of artificial
neural networks range from pattern recognition to the prediction of
financial markets. In particular, artificial neural networks are now a
well established technique in high energy physics, where they are used
for event reconstruction in particle detectors~\cite{netrev}.  Here we
will give a brief introduction to the specific type of artificial
neural networks (henceforth simply neural networks) which we shall use
for our analysis of structure functions. In particular, we will
describe multilayer feed--forward neural networks (or perceptrons),
and the algorithm used to train them (back-propagation learning).

\subsection{Multilayer neural networks}

\noindent

An artificial neural network consists of a set of interconnected units
(neurons).  The state or activation of a given $i$-neuron, $\xi_i$, is
a real number, determined as a function of the activation of the
neurons connected to it. Each pair of neurons $(i,j)$ is connected by
a synapsis, characterized by a real number $\omega_{ij}$ (weight). Note
that the weights need not be symmetric.  The activation of each neuron
is a function $g$ of the difference between a weighted average of
input from other neurons and a threshold $\theta_i$: \bea
\xi_i=g\left(\sum_{j}\omega_{ij}\xi_j- \theta_i \right)  .
\eea

The activation function $g$ is in general non-linear. The simplest
example of activation function $g(x)$ is the step function
$g(x)=\Theta(x)$, which produces binary activation only. However,
it turns out to be advantageous to use an activation
function with two distinct regimes, linear and
non--linear, such as the sigmoid \beq
g(x)\equiv\frac{1}{1+e^{-\beta x}}.
\label{sigdef}
\eeq 
This function
approaches the step function at large $\beta$; without loss of
generality we will take  $\beta=1$. The
sigmoid activation function has  a linear response when $x\approx 0$,
and it saturates  for large
positive or negative arguments. If  
weights and thresholds are  such that the
sigmoids work on the crossover between linear and saturation
regimes, the  neural network behaves
in a non-linear way. Thanks to this non--linear behaviour, the 
neural  network is  able to reproduce nontrivial functions.

We will in particular consider Rosenblatt's perceptrons, also known as
multilayer feed-forward neural networks~\cite{rosenblatt}.  These
networks are organized in ordered layers whose neurons only receive
input from a previous layer. For $L$ layers with $n_1,\ldots,n_L$
units respectively, the state of the multilayer neural network is
established by the recursive relations 
\beq
\xi_i^{(l)}=g\left(\sum_{j=1}^{n_{l-1}}\omega_{ij}^{(l-1)}\xi_j^{(l-1)}
- \theta_i^{(l)}\right);\quad i=1,\ldots,n_l,\quad l=2,\ldots,L,
\eeq 
where $\boldsymbol{\xi}^{(l)}$ represents the state of the
neurons in the $l^{th}$ layer, $\omega_{ij}^{(l)}$ the weights between
units in the $(l-1)^{th}$ and the $l^{th}$ layers, and
$\theta_i^{(l)}$ the threshold of the $i^{th}$ unit in the $l^{th}$
layer. The input is the vector $\boldsymbol{\xi}^{(1)}$ and the
output the vector $\boldsymbol{\xi}^{(L)}$.

\begin{figure}[t]
\begin{center}
\epsfig{width=0.7\textwidth,figure=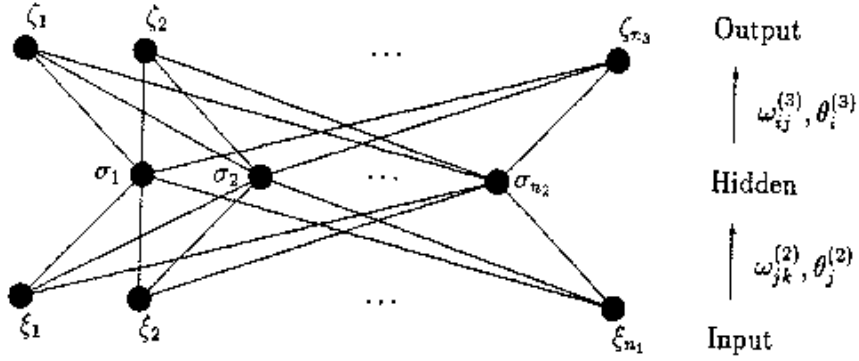}
\end{center}
\caption{A three-layer feed-forward neural network consisting of
input, hidden and output layers.}
\label{hidden}
\end{figure}
Multilayer feed-forward neural networks can be viewed as functions $F$
from $\mathbb{R}^{n_1}\rightarrow \mathbb{R}^{n_L}$, parametrized by
weights, thresholds and activation function, \beq
\boldsymbol{\xi}^{(L)}= F\left[ \boldsymbol{\xi}^{(1)}; \{
w_{ij}^{(l)} \} , \{\theta_i^{(l)}\}; g\right]  .
\label{acdef}
\eeq 
For given activation function, the 
parameters can be tuned in such a way that the neural
network reproduces any continuous function. 
The behaviour of a neural network is
determined by the joint behaviour of all its connections and
thresholds, and it can thus be built to be redundant, in the sense
that modifying, adding or removing a neuron 
has little impact on
the final output.  Because of these reasons, neural networks can be
considered to be robust, unbiased universal approximants.

\subsection{Learning process for neural networks}

\noindent
The usefulness of neural networks is due to the availability of a
training algorithm. This algorithm allows one to select the values of
weights and thresholds such that the neural network reproduces a given
set of input--output data (or patterns).  This procedure is called
`learning' since, unlike a standard fitting procedure, there is no
need to know in advance the underlying rule which describes the
data. Rather, the neural network generalizes the examples used to
train it.

The learning algorithm which we shall use is known as supervised
training by back-pro\-pa\-ga\-tion. Consider a set of input-output patterns
$({\bf x},{\bf z})$ which we want the neural network to learn. The
state of the neural network is generically given by 
\beq
\xi_i^{(l)}=g(h_i^{(l)}),\quad\quad h_i^{(l)}=\sum_{j=1}^{n_{l-1}}
\omega_{ij}^{(l-1)}\,\xi_j^{(l-1)}-\theta_i^{(l)}; \qquad
i=1,\dots,n_l, \quad l=1,\dots
,L .  \eeq 

Input--output patterns correspond to pairs of states of
the first and last layers, which we shall denote as 
\beq
 \matrix{{\bf x} &=& \boldsymbol{\xi}^{(1)}\,, \cr {\bf o}({\bf x})
&=& \boldsymbol{\xi}^{(L)}.}   \eeq 
The goal of the training is to
learn from a given set of data patterns, which consist of associations
of a given input with a desired output.  For any given values of the
weights and thresholds, define the error function \beq
E[{\omega,\theta}]\equiv\frac{1}{2}\sum_{A=1}^{n_p}\sum_{i=1}^{n_L}
(o_i({\bf x}^{A})-z_i^{A})^2\,,
\label{energy}
\eeq where $n_p$ is the number of data, \ie, the number of
input--output patterns. The error Eq.~(\ref{energy}) is the quadratic
deviation between the actual and the desired output of the network,
measured over the training set.

Given a fixed activation function $g$, Eq.~(\ref{acdef}), the error
function is a function of the weights and thresholds. It can be
minimized by looking for the direction of steepest descent in the
space of weights and thresholds, and modifying the parameters in that
direction: \bea \label{naivestpdesc}
\delta\omega_{ij}^{(l)} &=& -\eta\frac{\partial
E}{\partial\omega_{ij}^{(l)}}\,, \nonumber \\ \\ \nonumber
\delta\theta_{i}^{(l)} &=& - \eta\frac{\partial
E}{\partial\theta_{i}^{(l)}},
\eea where $\eta$ fixes the rate of descent, \ie~the `learning rate'.

The nontrivial result on which training is based is that it can be
shown that the steepest descent direction is given by the following
recursive expression: 
\bea \frac{\partial
E}{\partial\omega_{ij}^{(l)}} &=& \sum_{A=1}^{n_p}
\Delta_i^{(l)A}\xi_j^{(l-1)A};\qquad
i=1,\ldots,n_l,\quad j=1,\ldots,n_{l-1}\,, \nonumber \\ \frac{\partial
E}{\partial\theta_{i}^{(l)}} &=& -\sum_{A=1}^{n_p}
\Delta_i^{(l)A},\qquad i=1,\ldots,n_l,
\label{delpars}
\eea where the information on the goodness of fit is fed to the last
layer through \beq \Delta_i^{(L)A}= g'\l(h_i^{(L)A}\r)[o_i({\bf
x}^{A})-z_i^{A}]\,
\label{deldef}
\eeq and then it is back-propagated to the rest of the network by \beq
\Delta_j^{(l-1)A}=g'\l(h_j^{(l-1)A}\r) \sum_{i=1}^{n_l}
\Delta_i^{(l)A}\omega_{ij}^{(l)}\,.
\label{backprop}
\eeq

The back-propagation algorithm then consists of the following steps:
\begin{enumerate}
\item initialize all the weights and thresholds randomly, and choose a
small value for the learning rate $\eta$;
\item run a pattern ${\bf x}^A$ of the training set and store the
activations of all the units;
\item calculate $\Delta_i^{(L)A}$ Eq.~(\ref{deldef}), and then
back-propagate the error using Eq.~(\ref{backprop});
\item compute the contributions to $\delta\omega_{ij}^{(l)}$ and
$\delta\theta_{i}^{(l)}$ substituting Eq.~(\ref{delpars}) in
Eq.~(\ref{naivestpdesc});
\item update weights and thresholds.
\end{enumerate}
The procedure is iterated until a suitable convergence criterion is
fulfilled: for instance, if the value of the error function
Eq.~(\ref{energy}) stops decreasing.  The back--propagation can be
performed in ``batched'' or ``on--line'' modes. In the former case,
weights and thresholds are only updated after all patterns have been
presented to the network, as indicated in Eq.~(\ref{delpars}). In the
latter case, the update is performed each time a new pattern is
processed.  Generally, on-line processing leads to faster learning,
is less prone to getting trapped into local minima,
and it is thus preferable if the number of data is reasonably small
(\ie~$\lsim 10^3$).

It is clear that back-propagation seeks minima of the er\-ror
func\-tion given in Eq.~(\ref{energy}), but it cannot ensure that the
global minimum is found. Several modifications have been proposed to
 improve the algorithm so that local minima are avoided.  One of the
most successful, simple and commonly used variants is the introduction
of a `momentum' term. This means that Eq.~(\ref{naivestpdesc}) is
replaced by 
\bea\label{momterm}
 \delta\omega_{ij}^{(l)} &=& -\eta\frac{\partial
E}{\partial\omega_{ij}^{(l)}} +\alpha \
\delta\omega_{ij}^{(l)}(\rm{last}) \nonumber \\ 
 \delta\theta_{i}^{(l)} &=& -\eta\frac{\partial E}{\partial\theta_{i}^{(l)}} +\alpha \ \delta\theta_{i}^{(l)}(\rm{last}) 
\eea
where ``last'' denotes the values of the $\delta\omega_{ij}^{(l)}$ and
$\delta\theta_{i}^{(l)}$ used in the previous updating of the weights
and thresholds. The parameter $\alpha$ (`momentum')
must  be a positive number smaller than 1.

Also, it may be convenient to choose a more general form of the
error function, such as
\beq
E[{\omega,\theta}]\equiv\sum_{A,B=1}^{n_p}\sum_{i}^{n_L}
(o_i({\bf x}^{A})-z_i^{A})V_{AB}(o_i({\bf x}^{B})-z_i^{B})\,,
\label{genenergy}
\eeq
where $V_{AB}$ is a $n_p\times n_p$  matrix which can accommodate
weights and correlations between data, such as, typically, the
inverse of the covariance matrix. If the matrix $V_{AB}$ is
diagonal, the back--propagation algorithm is essentially unchanged, up
to a rescaling of $\Delta_i^{(L)A}$ Eq.~(\ref{deldef}) by the
corresponding matrix element $V_{AA}$. However, if the matrix $V_{AB}$
is not diagonal (\ie~the data are correlated), Eq.~(\ref{deldef})
must be substituted by a non--local expression which involves the sum
over all correlated data. Clearly, in this case  batched training is
preferable, so that all correlated patterns are shown to the net
before the weights and thresholds are updated.

\section{Neural structure functions}

\subsection{General Strategy}

We now describe in detail the strategy which we shall follow in order
to construct a neural parametrization of structure functions (see Figure~3).
As discussed in Sect.~2.2, 
the basic idea of our approach consists of constructing a
representation of the probability measure in the space of structure
functions. We do this in two steps: 
first, we generate a sampling of this measure based on the 
available data, and then we interpolate between these points using
neural networks. The final set of neural networks is the sought--for
representation of the probability measure. 

The first step consists of generating 
a Monte Carlo set of `pseudo--data', \ie~
$N_{rep}$ replicas of the original set of
$N_{dat}$ data points:
\beq
F^{(art)(k)}_i; \qquad k=1,\dots,N_{rep},\quad  i=1,\dots,N_{dat},
\label{replicas}
\eeq 
where the subscript $i$ is a shorthand for  $(x_i,Q_i^2)$.
The $N_{rep}$ sets of $N_{dat}$ points are 
distributed according to an $N_{dat}$--dimensional 
multigaussian distribution about the
original points, with expectation values equal to the central
experimental values, and error and covariance equal to the
corresponding experimental quantities. Because~\cite{dago,cowan}
the distribution of the experimental data coincides (for a flat prior)
with the probability distribution of the value of the structure
function $F_2$ at the points where it has been measured, this Monte
Carlo set gives a sampling of the probability measure at those points.
We can then generate arbitrarily many sets of pseudo--data: 
the number of sets $N_{rep}$ can be chosen to be large enough
that all relevant properties of the probability distribution of the
data (such as errors and correlations) 
are correctly reproduced by the given sample. 
We will perform a number of tests in order to determine the optimal
value of $N_{rep}$, which will be discussed in Sect.~4.2.

The second step consists of
training $N_{rep}$  neural networks, one for   each 
set of $N_{dat}$ pseudo--data.
The key issue here is the choice of a suitable error function
Eq.~(\ref{genenergy}). A possibility would be to use the simplest
unweighted form Eq.~(\ref{energy}):
\beq
E^{(k)}[\omega,\theta]=
\sum_{i=1}^{N_{dat}} {\l( F^{(art)(k)}_i-F^{(net)(k)}_i \r)^2},
\label{simpenergy} 
\eeq
where $F_i^{(net)(k)}$ is  the output of the neural network for  the
input values of  $(x,\,Q^2)$ corresponding to the $i^{th}$ data point.
 Then, with suitable choices of  architecture and training,
we can
achieve a value of the error such that $E/N_{dat}$ is much smaller
than the typical experimental uncertainty on each point.
In other words, the
trained net will then  go on top of each
point, on the scale of experimental errors. 
Therefore, it will reproduce the sampling of the probability
distribution at the measured points, and interpolate between them, thus
achieving our goal. In fact, this happens even if  $E/N_{dat}$ is not
significantly
smaller than the experimental uncertainty, provided only that for each
point $i$ the shifts $\delta_i^{(k)}\equiv F^{(net)(k)}_i-F^{(art)(k)}_i$
are distributed as uncorrelated random variables with zero mean as $k$
runs over replicas.
 Then, it is easy to see that the distribution of
$F^{(net)(k)}_i$ has the same average, variance, and point--to--point
correlations  as the distribution
of $F^{(art)(k)}_i$.

This is however not an optimal solution in that it does not fully
exploit the available physical information. To understand this, assume
the value of the structure function at the same $(x,\,Q^2)$ point is
measured by two different experiments. If the measurements are
independent, they are uncorrelated. However, since they determine the
same physical observable, they cannot be treated as independent, and
they must in fact be
combined to give a single determination,
with an error that it is smaller than either of the two measurements.
Now, it is clear that if the structure function is measured at 
two points which are very close in $(x,\, Q^2)$, then the value of the
structure function at these points are likewise not independent: 
even though we do not know the functional form of the structure
function, obvious QCD arguments imply that such a functional form 
exists, and it is continuous. Hence, by continuity, two neighbouring
points cannot be treated as independent samplings of a probability
distribution, and should be combined.
This combined distribution has a variance which is in general
smaller than its sampling at fixed
points.

This reduction
in variance is automatically achieved when fitting a given function
to the data:
if we fit a
function to each set of pseudo--data, then
as the pseudo--data span their multigaussian space, the best--fit
values of the parameters which parametrize the function vary over
their own space. As the number of data points is increased, the
variance of these parameters decreases. 
For instance, assuming for the
sake of argument a multilinear law with $k$ parameters, with $k$
data points all parameters can be determined with an error which can
be found by error propagation, but with $N_{dat}>k$ points this error
is reduced by a factor $1/\sqrt{N_{dat}-k}$. 
More in general, if the number of data exceeds the number of
parameters, the
error on the parameters will be reduced by a factor which  is 
bounded by the so--called RCF inequality~\cite{cowan}; if there exists an estimate
of the parameters which saturates this bound (as in the multilinear
case), then it can be found by  maximum--likelihood. 

We are thus 
led to think that the optimal strategy consists of choosing as an error
function the log-likelihood function for the given data. In this case, the 
matrix $V_{AB}$ in Eq.~(\ref{genenergy}) is identified with
the experimental covariance matrix, so the error function is
\beq
E^{(k)}[\omega,\theta]=
\sum_{i,j=1}^{N_{dat}} {\l( F^{(art)(k)}_i-F^{(net)(k)}_i \r)
\mathrm{cov}_{ij}^{-1}\l( F^{(art)(k)}_j-F^{(net)(k)}_j \r)},
\label{fullenergy} 
\eeq
where $\mathrm{cov}_{ij}$ 
is the covariance matrix defined in Eq.~(\ref{covdef}).
In such case, if we assume that after training the neural network can
capture --- at least locally ---
the underlying functional form, then a maximum--likelihood
determination of  its parameters leads to a distribution of neural
networks with a smaller uncertainty than that based on the
minimization of the naive error function Eq.~(\ref{simpenergy}).

In practice, however, this choice turns out not to be viable. Indeed,
as discussed in Sect.~3.2, the back--propagation algorithm with the
error function Eq.~(\ref{fullenergy}) is non--local, and has poor convergence
properties, especially when the number of data is large, as in our
case. Note that this difficulty is related to the number of correlated
data points, and thus it persists even if the number of
correlated systematics is actually small, despite the fact that in
such  case the inversion of the correlation matrix is not
a problem~\cite{cteq6}.  

A strategy that is both physically effective and numerically viable
consists instead of choosing as an error function the log--likelihood
calculated from uncorrelated statistical errors, namely
\beq
E^{(k)}[\omega,\theta]=
\sum_{i=1}^{N_{dat}} \frac{\l( F^{(art)(k)}_i-F^{(net)(k)}_i \r)^2}
{{\sigma^{(exp)}_{i,s}}^2}
\label{diaenergy}\, ,
\eeq
where 
$\sigma^{(exp)}_{i,s}$ is the statistical error on the $i^{th}$ data point.
Because the error function is now diagonal, use of the back--propagation
algorithm is no longer problematic. In order to
understand the effect of minimizing the error function
Eq.~(\ref{diaenergy}),
 write the
probability distribution of the $i^{th}$ data point as $k$ runs over
replicas as
\beq
F_i^{(art)(k)}= F_i^{(exp)}+ \sum_{p=1}^{N_{sys}} r_{i,p}^{(k)} \sigma_{i,p}+ 
r_{i,s}^{(k)} \sigma_{i,s},
\label{probdis}
\eeq
where $F_i^{(exp)}$ is the experimental value, 
all $r^{(k)}$ are distributed in $k$ as univariate gaussian
random numbers with
zero mean, $\sigma_s$ is the
statistical error, and $\sigma_p$ are the $N_{sys}$ systematics. Whereas
there are $N_{dat}$ independent $r_{i,s}$, the number of independent
variables $r_{i,p}$ for each $p$ is smaller, in that  there is
one single random variable $r_{i,p}$ for all correlated data points. The
maximum--likelihood fit of the error function Eq.~(\ref{diaenergy})
to a set of data distributed according  to Eq.~(\ref{probdis})
determines the values of $F^{(net)}_i$ which best fit the distribution of
\beq
F_i^{(sys)(k)}\equiv F_i^{(exp)}+ \sum_{p=1}^{N_{sys}} r_{i,p}^{(k)} \sigma_{i,p}. 
\label{fsydef}
\eeq 
In this case, the distribution of best--fit functional neural networks
as $k$ runs over replicas will have the smaller uncertainty which is
obtained by maximum--likelihood combination of the statistical
errors. However, it will still reproduce the (unimproved) systematic
errors
as $k$ runs over the Monte Carlo sample. The distribution of
systematic errors will remain unbiased, provided only
that for each point $i$ the distribution of
best--fit values of $F^{(net)(k)}_i$  about 
$F^{(art)(k)}_i$ as  $k$ runs over replicas is such that
$F^{(net)(k)}_i$ is uncorrelated  to $r_{i,p}^{(k)}$.
In other words, the best--fit nets provides a probability distribution which
optimally combines statistical errors, but reproduces the systematic
errors of the Monte Carlo sample.

Just as in the first step, also in this second step of the procedure
it will be necessary to verify whether the number of replicas is large
enough that the statistical properties of the distributions are
properly reproduced. Furthermore, it will be necessary to verify that
the network fitting procedure did not introduce any
bias, and specifically that if the statistical variance is reduced
this is due to having combined data points, and not to a bias of the fitting
procedure. To this purpose, we will develop a number of statistical
tools which will be discussed in Sect.~5.1 when presenting our results.

At the end of our procedure, we obtain a set of $N_{rep}$
trained neural networks which provides us with a representation of the
probability measure in the space of structure functions, and thus it allows us
to determine any observable by averaging with this measure,~Eq.~(\ref{funave}). 
In particular, estimators for expectation values, errors
and correlations are 
\bea
\Big\lan F^{(net)}_i\Big\ran_{rep}=\frac{1}{N_{rep}}\sum_{k=1}^{N_{rep}}
F_i^{(net)(k)}
\label{valesdef}\\
\sigma^{(net)}_i=\sqrt{\Big\lan \l( F^{(net)}_i\r)^2\Big\ran_{rep}-\Big\lan
F^{(net)}_i\Big\ran_{rep}^2}
\label{sigesdef}\\
\rho^{(net)}_{ij}=\frac{\Big\lan F^{(net)}_i\,F^{(net)}_j\Big\ran_{\ngen}
- \Big\lan F^{(net)}_i\Big\ran_{\ngen}
\Big\lan
F^{(net)}_j\Big\ran_{\ngen}}{\sigma^{(net)}_i\,\sigma^{(net)}_j}\, ,
\label{coresdef}\eea
where $i$, $j$ denote two pairs of values $(x_i,Q^2_i)$, $(x_j, Q^2_j)$ 
(not necessarily in the original data set).
More in general, any functional average, defined by
Eq.~(\ref{funave}),
can be estimated by 
\beq
\blan {\cal F}
\left[F_2(x,Q^2)\right]\bran=\frac{1}{N_{rep}}\sum_{k=1}^{N_{rep}} 
{\cal F} \left[{F_2}^{(net)(k)}(x,Q^2)\right].
\label{discfunave}
\eeq

\begin{figure}[t]
\begin{center}
\epsfig{width=1.0\textwidth,figure=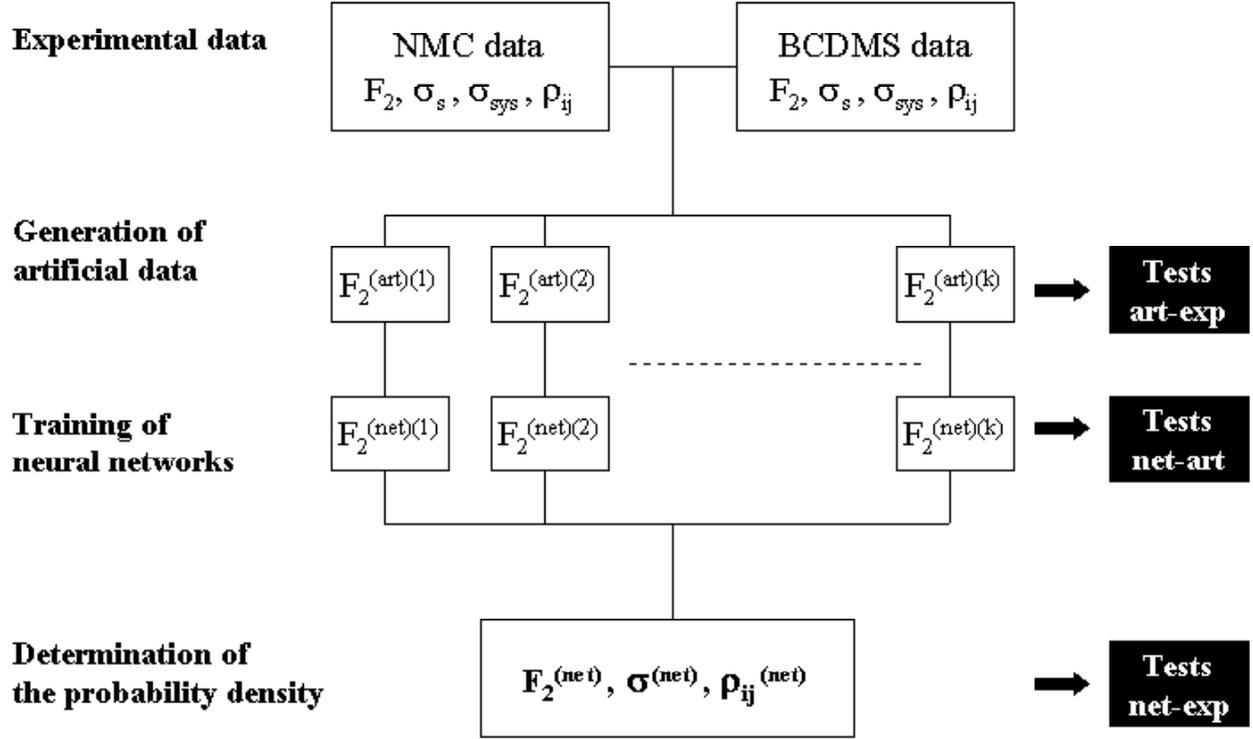}  
\end{center}
\begin{center}
\caption{Flow chart for the construction of  the parametrization of
structure functions}
\label{fig:flowchart}
\end{center}
\end{figure}

\subsection{Generation of artificial data}
\noindent

The Monte Carlo set of pseudo--data can be generated by using equations
of the form of~(\ref{probdis}). In particular, for the NMC experiment
we have
\bea
F_i^{(art)\,(k)}&=& (1+r_7^{(k)}\,\sigma_{N}) \Bigg[ F_i^{(exp)}  \\ \nonumber 
&+& \frac{r_{i,1}^{(k)}\,E_i+r_{i,2}^{(k)}\,E'_i+r_{i,3}^{(k)}\,AC_i+
r_{i,4}^{(k)}\,RC_i+r_{i,5}^{(k)}\,RE_i}{100}F_i^{(exp)}+
r_{i,s}^{(k)}\,\sigma_{i,s}\Bigg]\,,
\label{gennmc}
\eea
where each  $r_{i,p}^{(k)}$ is distributed as an
univariate gaussian random number
with zero mean  over the replica sample
(\ie, as $k$ varies), $F_i^{(exp)}$ is the central experimental value, and
the various sources of systematics were discussed in Sect.~2.1.1. 
Since statistical errors are uncorrelated, there is an independent $r_{i,s}$
for each data point. 
Correlated systematics are correctly reproduced by
taking the same  $r_3$ and $r_5$  for all data corresponding to all
energies and targets, the same $r_4$ for all targets, and
the same $r_1$,
$r_2$ and $r_7$ for all targets.

Likewise, for  BCDMS we have
\bea
F_i^{(art)\,(k)}&=& (1+r_5^{(k)}\,\sigma_{N}) \sqrt{1+r_{i,6}^{(k)}\,\sigma_{N_t}}
\sqrt{1+r_{i,7}^{(k)}\,\sigma_{N_b}}
\Bigg[ F_i^{(exp)}  
\label{genbcdms}
\\ \nonumber 
&+&
\frac{r_{i,1}^{(k)}\,f_b+r_{i,2}^{(k)}\,f_{i,s}+r_{i,3}^{(k)}\,
f_{i,r}}{100}F_i^{(exp)}+
r_{i,s}^{(k)}\,\sigma_{s}^i\Bigg]\,,
\eea
where now $r_1$,$r_2$,$r_3$ and $r_5$ take the same value for all data
(all energies and targets), $r_6$ is fixed for a given target, $r_7$
is fixed for given beam energy.

In order to determine a suitable value of $N_{rep}$, we now compare
expectation values, variance and correlations of the pseudo--data set
with the corresponding input experimental values, as a function of the
number of replicas.  A graphical comparison is shown in Figure.~4,
where we display scatter plots of the central values, errors and
correlation coefficients for samples of 10, 100 and 1000 replicas, for
the proton, deuteron, and nonsinglet structure functions.
\begin{figure}[t]
\begin{center}
\epsfig{width=0.75\textwidth,figure=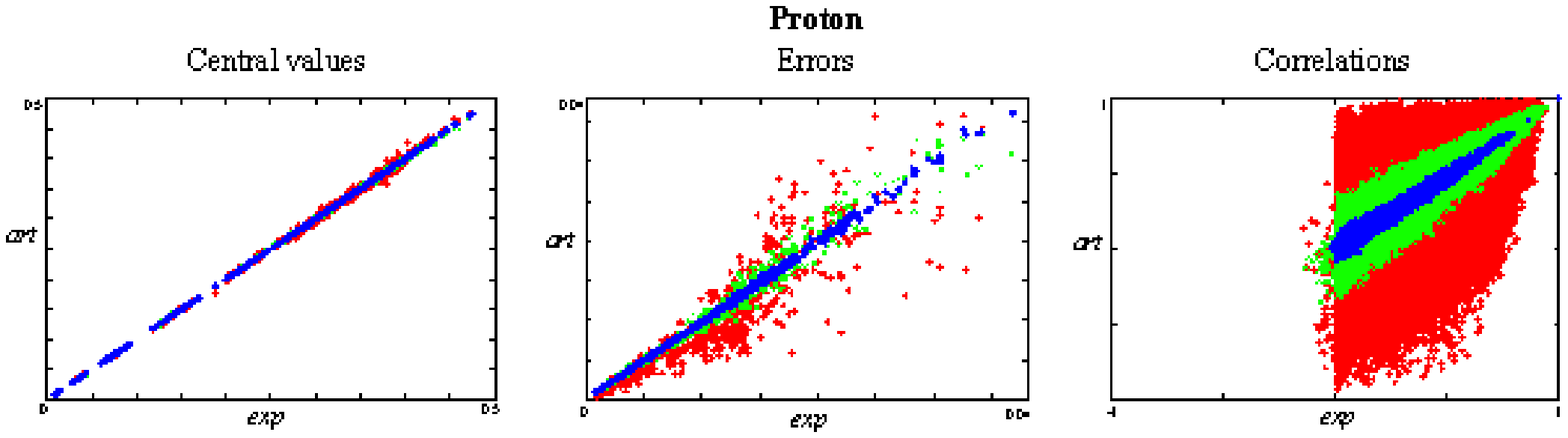}  
\epsfig{width=0.75\textwidth,figure=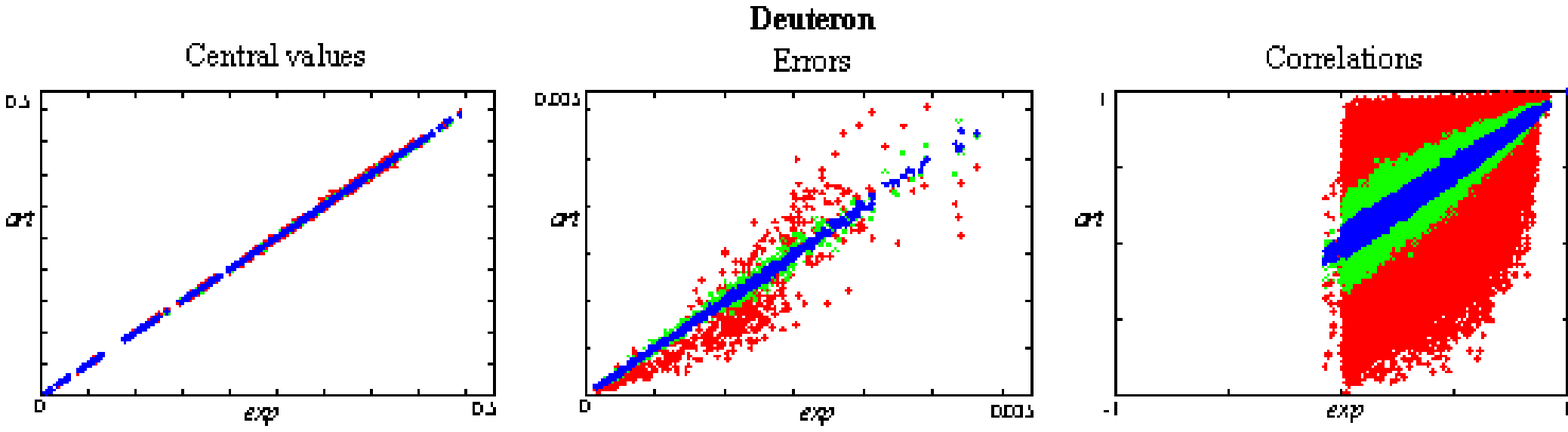}  
\epsfig{width=0.75\textwidth,figure=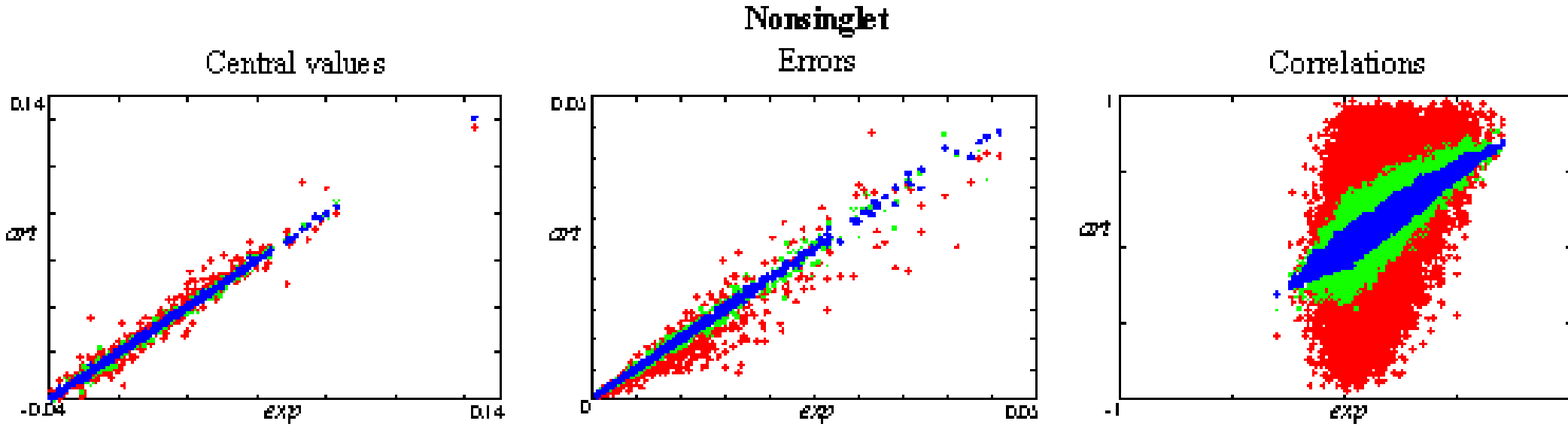}  
\end{center}
\caption{$\Big\lan F^{(art)}_i\Big\ran_{\ngen}$ vs. $\bar
F^{(exp)}_i$, $\Big\lan \sigma^{(art)}_i\Big\ran_{\ngen}$ vs. $\bar
\sigma^{(exp)}_i$
 and
$\Big\lan\rho_{ij}^{(art)}\Big\ran_{\ngen}$ 
vs. $\rho_{ij}^{(exp)}$ with $N_{\ngen}=$~10
(red), 
100 (green)  and 1000
(blue) replicas.}
\label{fig:genexp}
\end{figure}

\begin{table}[t]  
\def\ngen{rep}

\begin{center}  
\begin{tabular}{cccc} 
\multicolumn{4}{c}{$F_2^p$}\\ 
\hline  
$N_{\ngen}$ & 10 & 100 & 1000 \\
\hline  
$\Big\lan V\l[\Big\lan F^{(art)} \Big\ran_{\ngen}\r]\Big\ran_{dat}$ 
  & $1.8\times 10^{-5}$ & $1.7\times 10^{-6}$ & $1.3\times 10^{-7}$\\
$\Big\lan PE\l[\Big\lan F^{(art)} \Big\ran_{\ngen}\r]\Big\ran_{dat}$ 
  & 1.3\% & 0.4\% & 0.1\% \\
$r [F^{(art)}]$ 
  &  0.99949 & 0.99995 & 0.99999 \\
\hline
$\Big\lan V[\sigma^{(art)} ]\Big\ran_{dat}$ 
  & $8.0\times 10^{-5}$ & $2.4\times 10^{-5}$ & $7.0\times 10^{-6}$ \\
$\Big\lan PE[\sigma^{(art)} ]\Big\ran_{dat}$ 
  & 40\% & 11\% & 4\% \\
$\Big\lan \sigma^{(art)} \Big\ran_{dat}$ 
  & 0.0102 & 0.0114 &  0.0114\\
$r [\sigma^{(art)} ]$ 
  & 0.900 & 0.987 &  0.999\\
\hline
$\Big\lan V[\rho^{(art)}]\Big\ran_{dat}$ 
  & 0.0868 & 0.0056 &  0.0006\\
$\Big\lan \rho^{(art)} \Big\ran_{dat}$ 
  & 0.348 & 0.384 &  0.378\\
$r [\rho^{(art)}]$ 
  & 0.623 & 0.960 & 0.996 \\
\hline
$\Big\lan V[\mathrm{cov}^{(art)}]\Big\ran_{dat}$ 
  & $5.5\times 10^{-9}$ & $4.6\times 10^{-10}$ & $ 4.7\times 10^{-11}$ \\
$\Big\lan \mathrm{cov}^{(art)} \Big\ran_{dat}$ 
  & $3.4\times 10^{-5}$ & $3.8\times 10^{-5}$ & $ 3.8\times 10^{-5}$\\
$r [\mathrm{cov}^{(art)}]$ 
  & 0.519 & 0.906 & 0.987 \\
\hline
\end{tabular}
\end{center}
\caption{Comparison between experimental and generated artificial data for
the proton structure function. 
The experimental data have $\Big\lan \sigma^{(exp)}
\Big\ran_{dat}=0.0116$, $\Big\lan \rho^{(exp)} \Big\ran_{dat}=0.382$
and $\Big\lan \mathrm{cov}^{(exp)} \Big\ran_{dat}=3.8\times 10^{-5}$.}
\label{Tgenexpp}
\end{table}
\begin{table}[t]  
\begin{center}  
\begin{tabular}{cccc} 
\multicolumn{4}{c}{$F_2^d$}\\   
\hline
$N_{\ngen}$ & 10 & 100 & 1000 \\
\hline  
$\Big\lan V\l[\Big\lan F^{(art)} \Big\ran_{\ngen}\r]\Big\ran_{dat}$ 
  & $9.2\times 10^{-6}$ & $9.1\times 10^{-7}$ & $9.7\times 10^{-8}$\\
$\Big\lan PE\l[\Big\lan F^{(art)} \Big\ran_{\ngen}\r]\Big\ran_{dat}$ 
  & 1.1\% & 0.3\% & 0.1\% \\
$r [F^{(art)}]$ 
  & 0.99976  & 0.99998 & 0.99999 \\
\hline
$\Big\lan V[\sigma^{(art)} ]\Big\ran_{dat}$ 
  & $5.3\times 10^{-5}$ & $1.5\times 10^{-5}$ & $4.0\times 10^{-6}$ \\
$\Big\lan PE[\sigma^{(art)} ]\Big\ran_{dat}$ 
  & 39\% & 11\% & 3\% \\
$\Big\lan \sigma^{(art)} \Big\ran_{dat}$ 
  & 0.0095 & 0.0102 & 0.0102 \\
$r [\sigma^{(art)} ]$ 
  & 0.857 & 0.990 & 0.999 \\
\hline
$\Big\lan V[\rho^{(art)}]\Big\ran_{dat}$ 
  & 0.0923 & 0.0075 & 0.0007 \\
$\Big\lan \rho^{(art)} \Big\ran_{dat}$ 
  & 0.374 & 0.310 & 0.310 \\
$r [\rho^{(art)}]$ 
  & 0.641 & 0.934 & 0.993 \\
\hline
$\Big\lan V[\mathrm{cov}^{(art)}]\Big\ran_{dat}$ 
  & $2.6\times 10^{-9}$ & $2.5\times 10^{-10}$ & $ 2.3\times 10^{-11}$ \\
$\Big\lan \mathrm{cov}^{(art)} \Big\ran_{dat}$ 
  & $3.3\times 10^{-5}$ & $3.3\times 10^{-5}$ & $ 3.2\times 10^{-5}$\\
$r [\mathrm{cov}^{(art)}]$ 
  & 0.568 & 0.932 & 0.992 \\
\hline
\end{tabular}
\end{center}
\caption{Comparison between experimental and generated artificial data for
the deuteron structure function. 
The experimental data have $\Big\lan \sigma^{(exp)}
\Big\ran_{dat}=0.0102$, $\Big\lan \rho^{(exp)} \Big\ran_{dat}=0.313$
and $\Big\lan \mathrm{cov}^{(exp)} \Big\ran_{dat}=3.3\times 10^{-5}$.}
\label{Tgenexpd}
\end{table}
\begin{table}[t]  
\begin{center}  
\begin{tabular}{cccc} 
\multicolumn{4}{c}{$F_2^p-F_2^d$}\\   
\hline  
$N_{\ngen}$ & 10 & 100 & 1000 \\
\hline  
$\Big\lan V\l[\Big\lan F^{(art)} \Big\ran_{\ngen}\r]\Big\ran_{dat}$ 
  & $1.4\times 10^{-5}$ & $1.8\times 10^{-6}$ & $2.7\times 10^{-7}$\\
$\Big\lan PE\l[\Big\lan F^{(art)} \Big\ran_{\ngen}\r]\Big\ran_{dat}$ 
  & 35\% & 11\% & 4\% \\
$r [F^{(art)}]$ 
  &  0.980 & 0.998 & 0.999 \\
\hline
$\Big\lan V[\sigma^{(art)} ]\Big\ran_{dat}$ 
  & $6.6\times 10^{-5}$ & $2.19\times 10^{-5}$ & $7.8\times 10^{-6}$ \\
$\Big\lan PE[\sigma^{(art)} ]\Big\ran_{dat}$ 
  & 35\% & 12\% & 4\% \\
$\Big\lan \sigma^{(art)} \Big\ran_{dat}$ 
  & 0.0101 & 0.0114 & 0.0114 \\
$r [\sigma^{(art)} ]$ 
  & 0.927 & 0.991 & 0.999 \\
\hline
$\Big\lan V[\rho^{(art)}]\Big\ran_{dat}$ 
  & 0.1133 & 0.0094 & 0.0010 \\
$\Big\lan \rho^{(art)} \Big\ran_{dat}$ 
  & 0.1112 & 0.0990 & 0.0946 \\
$r [\rho^{(art)}]$ 
  & 0.405 & 0.816 & 0.971 \\
\hline
$\Big\lan V[\mathrm{cov}^{(art)}]\Big\ran_{dat}$ 
  & $5.1\times 10^{-9}$ & $5.8\times 10^{-10}$ & $ 6.0\times 10^{-11}$ \\
$\Big\lan \mathrm{cov}^{(art)} \Big\ran_{dat}$ 
  & $7.3\times 10^{-6}$ & $9.0\times 10^{-6}$ & $ 8.7\times 10^{-6}$\\
$r [\mathrm{cov}^{(art)}]$ 
  & 0.346 & 0.791 & 0.972 \\
\hline
\end{tabular}
\end{center}
\caption{Comparison between experimental and generated artificial data for
the nonsinglet structure function. 
The experimental data have $\Big\lan \sigma^{(exp)}
\Big\ran_{dat}=0.0114$, $\Big\lan \rho^{(exp)} \Big\ran_{dat}=0.090$
and $\Big\lan \mathrm{cov}^{(exp)} \Big\ran_{dat}=8.4\times 10^{-6}$.}
\label{TgenexpNS}
\end{table}
A more quantitative comparison can be performed defining the 
following quantities
\begin{itemize}
\item Average over the number of replicas for each experimental point $i$
(\ie~a pair of values of $x$ and $Q^2$)
\beq
\Big\lan F^{(art)}_i\Big\ran_{\ngen}=\frac{1}{N_{\ngen}}\sum_{k=1}^{N_{\ngen}}
F_i^{(art)(k)}\,.
\label{ravedef}
\eeq
Variance of $\lan F^{(art)}_i\ran_{\ngen}$ 
\beq
\sigma^{(art)}_i=\sqrt{\Big\lan \l( F^{(art)}_i\r)^2\Big\ran_{\ngen}
-\Big\lan F_i^{(art)}\Big\ran_{\ngen}^2}\,.
\label{rvardef}
\eeq
Correlation and covariance of two points
\bea
\rho^{(art)}_{ij}&=&\frac{\Big\lan F^{(art)}_i\,F^{(art)}_j\Big\ran_{\ngen} - \Big\lan F^{(art)}_i\Big\ran_{\ngen}
\Big\lan
F^{(art)}_j\Big\ran_{\ngen}}{\sigma^{(art)}_i\,\sigma^{(art)}_j},\\
\mathrm{cov}^{(art)}_{ij}&=&\rho^{(art)}_{ij} \sigma^{(art)}_i\sigma^{(art)}_j.
\label{rcordef}
\eea
These quantities provide the estimators  of the experimental values, errors,
and correlation which one extracts from the sample of artificial data.
\item Mean variance and mean percentage error on central values over 
the number of points $N_{dat}$ 
\bea
\Big\lan V\l[\Big\lan F^{(art)} \Big\ran_{\ngen}\r]\Big\ran_{dat} &=& 
\frac{1}{N_{dat}} \sum_{i=1}^{N_{dat}}\l(\Big\lan F^{(art)}_i\Big\ran_{\ngen}-F^{(exp)}_i\r)^2 \\
\Big\lan PE\l[\Big\lan F^{(art)} \Big\ran_{\ngen}\r]\Big\ran_{dat}
&=& 
\frac{1}{N_{dat}} \sum_{i=1}^{N_{dat}}\l|\frac{\Big\lan F^{(art)}_i\Big\ran_{\ngen}-F^{(exp)}_i}{F^{(exp)}_i}\r|\,. 
\eea
We can similarly define $\Big\lan V[\sigma^{(art)}]\Big\ran_{dat}$, 
$\Big\lan PE[\sigma^{(art)}]\Big\ran_{dat}$,
$\Big\lan V[\rho^{(art)}]\Big\ran_{dat}$
and $\Big\lan V[\mathrm{cov}^{(art)}]\Big\ran_{dat}$. 
These estimators 
indicate how close the averages over generated data are to the
experimental values. 
\item Scatter correlation: 
\beq
r[F^{(art)}]=\frac{\Big\lan F^{(exp)}\,
 \Big\lan F^{(art)}\Big\ran_{\ngen}  
\Big\ran_{dat} - \Big\lan F^{(exp)}\Big\ran_{dat}
\Big\lan \Big\lan F^{(art)}\Big\ran_{\ngen}  
\Big\ran_{dat}}{\sigma_s^{(exp)}\,\sigma_s^{(art)}}\,.
\label{sccdef}
\eeq
where the scatter variances are defined as
\bea\sigma_s^{(exp)}&=& \blan (F^{(exp)})^2\bran_{dat}-\left(\blan
F^{(exp)}\bran_{dat}\right)^2 
\nonumber\\
\sigma_s^{(art)}&=& \blan \left(\blan F^{(art)}\bran_{rep}\right)^2\bran_{dat}-
\left(\blan\blan F^{(art)}\bran_{rep}\bran_{dat}\right)^2.
\label{scsdef}
\eea
Similarly we define $r[\sigma^{(art)}]$, 
$r[\rho^{(art)}]$ and
$r[\mathrm{cov}^{(art)}]$. The scatter correlation indicates
the size of the
spread of data around a straight line 
in the scatter plots of Figure~\ref{fig:genexp}.
Specifically, $r[\sigma^{(net)}]=1$
implies that $\langle \sigma^{(net)}_i\rangle$  is proportional to
$\sigma^{(exp)}_i$
(and similarly for  $r[F^{(net)}]$ etc.).
 Note that the scatter
correlation and scatter variance are not related to the variance and
correlation Eqs.~(\ref{rvardef}-\ref{rcordef}) or averages thereof.
\item Average variance:
\beq
\blan \sigma^{(art)}\bran_{dat}=\frac{1}{N_{dat}}\sum_{i=1}^{N_{dat}}
\sigma^{(art)}_i. 
\label{avvardef}
\eeq
Similarly we can define $\blan \rho^{(art)}\bran_{dat}$ and
$\blan \mathrm{cov}^{(art)}\bran_{dat}$. Analogously, we can define
the corresponding experimental quantities 
\beq
\blan
\sigma^{(exp)}\bran_{dat}=\frac{1}{N_{dat}}\sum_{i=1}^{N_{dat}}\sigma^{(exp)}_i
\label{expave}
\eeq
where $\sigma^{(exp)}_i$ is the experimental error on the $i^{th}$ data
point, an so forth. These quantities are
interesting because even if the scatter correlation $r$
Eq.~(\ref{sccdef}) is very close to one (so all points in the scatter
plots 
Figure~4 lie on a straight line), there could still be
a systematic bias in the estimators
Eqs.~(\ref{ravedef}-\ref{rcordef}). 
For example, if for all points $i$ the variance
$\sigma_i^{(art)}=\lambda \sigma_i^{(exp)}$, then
$r[\sigma^{(art)}]=1$, but $\blan \sigma^{(art)}\bran_{dat}=\lambda
\blan \sigma^{(exp)}\bran_{dat}$.
\end{itemize}

The values of these quantities for samples of 10, 100 and 1000
replicas and for the three structure functions under consideration are
shown in Tables~1--3. From these tables, it can be seen that
the scaling of the various quantities with $N_{rep}$ follows the standard
behaviour of Gaussian Monte Carlo samples~\cite{cowan}.
For instance, the variance on central values  should scale as
$1/N_{\ngen}$, while the variance on the errors should scale as
$1/\sqrt{N_{\ngen}}$. This implies that a smaller sample
is sufficient in order to achieve a given accuracy on means than it is
required to reach the same accuracy on errors. 
Also, because $V[\rho^{(art)}]=[1-(\rho^{(exp)})^2]^2/N_{rep}$,
the estimated correlation fluctuates more for small values of
$\rho^{(exp)}$, and thus the average correlation
$\langle\rho^{(art)}\rangle_{dat}$ tends to be larger than the
corresponding experimental value  $\langle\rho^{(exp)}\rangle_{dat}$.
Notice finally that a larger sample is necessary in order to achieve a
given percentage error on the nonsinglet than  on the proton and
deuteron. This is a consequence of the fact that the signal--to--noise
ratio is smaller for the nonsinglet: the proton and deuteron structure
functions are roughly of the same size, so the absolute  value of the
nonsinglet structure function is typically by about a factor~10 smaller.

Inspection of the tables shows that in order
to
reach average scatter correlations of 99\% and percentage accuracies
of a few percent  on  structure functions,  errors and correlations a
sample of about 1000 replicas is necessary. Notice in particular that
with 1000 replicas  the estimated correlation fluctuates about the
true value, rather than systematically overshooting it.

\subsection{Building and training Neural Networks}
\noindent

The choice of an optimal structure of the neural nets and of an
optimal learning strategy cannot follow fixed rules and must be
tailored to the specific problem. Here we summarize and motivate the
specific choices which we made for our fits.

{\it Number of layers} 

The optimal number of layers in a neural network depends on the complexity 
of the specific task  it
 should perform. It can be proved~\cite{netrev}
that any  continuous function, no matter
 how complex, can be represented by a multilayer
neural network with no more than three layers (input layer, 
a middle hidden layer, and output layer). However, in practice,  
using a single hidden layer may require a very large number of units
in it. Thus, it can be more useful to have two hidden
layers and a smaller number of neurons on each. 
We will only consider networks with two
hidden layers.

{\it Number of hidden units}

The number of hidden units needed to approximate a given function 
${\cal F}$ is related to the number of  terms which
are needed in an expansion
of ${\cal F}$ over the basis of  functions $g(x),\> g(g(x)),\> \dots$.
There exist several techniques to determine the optimal number
of units.
Here we have taken a pragmatic approach. First, we 
carry out all the learning with a
small number of units. Then, we restart
 the whole learning adding new units  one by
one, until  stability of the error function is reached. 
In this way, an optimal
architecture is found, which is large enough to reproduce faithfully
the training
patterns, but small enough that training is fast. The
final results which we will present are obtained
with  the architecture (4--5--3--1). The reason why there are four
input nodes will be discussed below.

{\it Activation function}

We have taken a sigmoid activation function Eq.~(\ref{sigdef}).
The sensitivity of the neural network can be  enhanced
by substituting the sigmoid activation function in
the last layer with the identity (linear response). This
avoids saturation of the last layer neurons and leaves
space for more sensitive responses.  For this reason, we
have adopted linear response in the last layer. With these choices,
and the  architecture discussed above, the
behaviour of the network is determined by the values of 47 free
parameters (38 weights and 9 thresholds).

{\it Rescaling}

If the activation $\xi_i $ of input (output)
 node $i$ is numerically large, in order for the activation function
 to be in the nonlinear response regime, the weights $\omega_{ij}$
 must be very small. However,
if the activation is  very
 large, the shifts Eq.~(\ref{deldef}) in the first stages of the
 learning process are also very large, thus
 leading to an erratic behaviour of the training
 process, unless the learning rate $\eta$ Eq.~(\ref{delpars}) is
 initially  taken
 to be very small and then changed along the training. 
In order to avoid this, it is
convenient to  rescale both the input and the output 
data in such a way that that typical activations are    
$\xi_i\sim{\cal O}(1)$ for all $i$.
In practice, we have 
rescaled all values of $(x, Q^2)$ and the structure function $F_2$ in
 such a way that they span the range $0.1-0.9$. This ensures
that activation functions are in the nonlinear regime when
the  absolute values of all the  weights settle to values close to one. 

{\it Input}

 We  have used as
input variables $x$, $Q^2$, $\log x$ and $\log Q^2$. The choice of taking 
$\log x$ and $\log Q^2$ along with $x$ and $Q^2$ is motivated by the
expectation that  these are the variables upon which $F_2$
naturally depends. Note that this is not 
a source of theoretical bias: if this expectation turns out to be
incorrect and these variables are useless, after training  
the neural network will simply disregard them. If, however, the
expectation is correct this choice will speed up the training.
We have checked that neural networks trained just with 
$x$ and $Q^2$ perform as well as the ones we use but need
longer training times. 

{\it Theoretical assumptions}

The only theoretical assumption on the  shape of $F_2(x,Q^2)$
produced by the neural nets is that it satisfies the kinematic bound 
$F_2(1,Q^2)=0$ for all $Q^2$. In general, a constraint can be enforced
by adding it to the error function Eq.~(\ref{energy}), so that
configurations that violate the constraint are 
unfavourably weighted. In our case, the constraint
 is local in $x$ and $Q^2$ so its implementation is straightfoward: 
it can  be
enforced by  including in the data set
a number of artificial data points which
satisfy the constraint, with suitably tuned error. More general
cosntraints, such as the momentum sum rule in a parton fit, would
require  a term which makes the error function nonlocal in $x$ and
$Q^2$, and would thus make the training computationally more intensive.

We have thus
added  to the data set
10 artificial points at $x=1$ with equally spaced values
of $Q^2$ on a linear scale between $20$ and $200$~GeV$^2$ (which
ensures that the bound is respected even when extrapolating beyond the
data region). 
The choice of the error on these points is very delicate:
if it is too small, the neural networks will spend a significant fraction of
their training time in learning these points.  In such case, 
the kinematic constraint $F_2(x=1,Q^2)=0$ is enforced with great
accuracy, but at the expense of the fit to experimental data. 
We have thus taken the error on these points at $x=1$ to be of
the same order of the smallest experimental error, namely, 
$10^{-3}$ for $F_2^p$ and $F_2^d$, and  $\sqrt{2}\times10^{-4}$ for
$F_2^p-F_2^d$.

{\it Training patterns}

Since the number of data is reasonably
small, we have adopted on--line training (see Sect.~3.2):
weights are updated after each pattern is
shown to the net.  If the experimental 
data were shown to the net in a fixed order, 
a bias could arise, \eg\ because of the  
regularity of some input patterns. Therefore, we have shown
the data to the network using each time a random permutation of their
order. Furthermore, the data are always shown
at least $10^{4}$ times, in order to allow compensations
of different variations. Because we use a pseudo--random number
generator to compute permutations with periodicity $6\times 10^6$ 
which is smaller
than a typical number of training cycles (see Sect.~5), there might
still be 
periodic  oscillations in behaviour of the best--fit network.
However, we have checked that 
these oscillations do not significantly affect the fit.

{\it Learning parameters}

The choice of the learning rate $\eta$ Eq.~(\ref{naivestpdesc}) is
 particularly important: large values of $\eta$ lead to an unstable
 training process, while small values lead to slow training which may
 get trapped in local minima. 
In practice, it is convenient to vary the value of $\eta$ during the
 training: first, one looks for the region of the global minimum with
 a larger value of $\eta$.
Once this region is located, the learning rate is reduced 
  so that the probability of jumping away is
small and it is then easier to deepen into the minimum. 
The  momentum term $\alpha$ Eq.~(\ref{momterm}) is
closely connected to the learning rate:  an increase of $\alpha$ implies an increase
of the effective  learning rate. The optimum value of $\alpha$ depends on the updating
procedure used. 

For our training, we have chosen to first minimize the error function
Eq.~(\ref{simpenergy}), and then switch to the minimization of
 Eq.~(\ref{diaenergy}): first we look for the rough location of the minimum, and then
we refine its search. The first set of training  (of the order of
$10^6$  cycles) is
performed with $\eta\sim 10^{-3}$, and the second  (of the order of
$ 10^8$  cycles) with
$\eta\sim 10^{-8}$ (see Sect.~5). We will always take  $\alpha=0.9$. 

{\it Convergence of the training procedure and parametrization bias}

\begin{figure}[t]
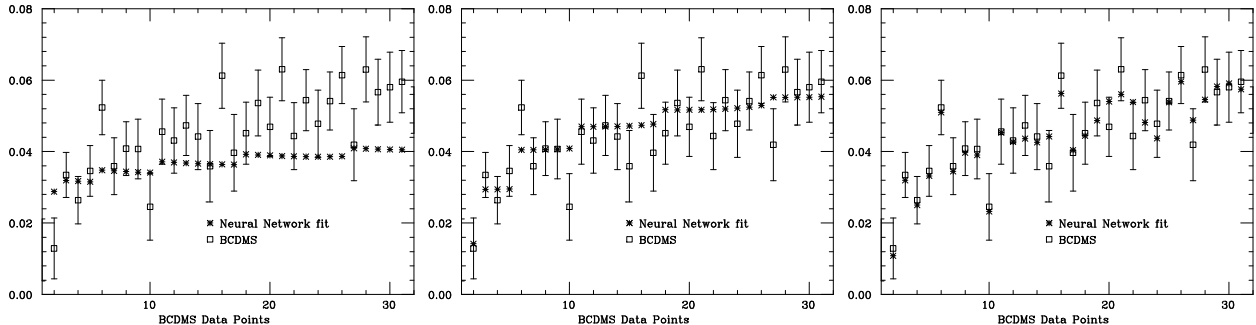

\begin{center}
\epsfig{width=0.32\textwidth,figure=st.ps}  
\epsfig{width=0.32\textwidth,figure=nt.ps}  
\epsfig{width=0.32\textwidth,figure=ol.ps}  
\end{center}
\caption{Fit of the nonsinglet structure function $F_2^p-F_2^d$ 
to a subset of BCDMS data points for increasing training lengths:
insufficient training (left);
< normal training (middle); overlearning (right).
The variable in abscissa is an arbitrary point number.}
\label{netcomp}
\end{figure}

As the training of the neural network progresses, the neural net
reproduces in a more and more detailed way the features of the given data set.
It follows that 
the problem of determining convergence of the training procedure is
tangled with the problem of avoiding a  possible fitting bias. Indeed, 
if the net
is too small or the training is too short, it will artificially smooth
out physical fluctuations in the data, whereas if the training is too
long, the net will also fit statistical noise (`overlearning'). 
Convergence must be therefore assessed on the basis of a
criterion which guarantees that the trained nets faithfully reproduce the
local functional form of the data, without fitting  statistical
fluctuations, but also without imposing a smoothing bias. 
This issue is illustrated in Figure~\ref{netcomp}, where we show the form of a
neural network with the
architecture and features discussed above, trained  to a small, arbitrarily
chosen subset of  30 BCDMS data points for the structure function
$F_2^p-F_2^d$, with
 increasingly long training cycles. 
With this  small data set, chosen for the sake of illustration,
the architecture of the nets is very
redundant, and overlearning is easily
achieved. 
 The first plot (very short
training) shows that the
neural network strongly correlates data points, but 
it does not quite reproduce the behavior of data. The
second plot corresponds to a
longer training, which leads  to a more or less ideal fit. In the last plot
(very long training)
the neural network follows the statistical fluctuations of individual
data points (overlearning). Note that the abscissa in this plot is an
arbitrary point number; jumps in the best--fit function are not
significant and simply reflect the fact that points of neighbouring
number might correspond to rather different values of $(x,Q^2)$.

Extreme overlearning as displayed in Figure~\ref{netcomp}
can be avoided by  using a number of parameters  in the 
neural network which, while still being largely redundant so as to
avoid parametrization bias,  is significantly smaller 
than  the number of points to be fitted. 
However, an accurate assessment of whether convergence
has been reached can only be obtained by means of statistical
indicators of the goodness of fit, tailored  to the particular
problem. These will be discussed in the next section. By means of
such indicators, one may determine an optimal training length, and then
subject all networks to training of  this fixed length. This is more
convenient than  stopping the training on the basis of a convergence
criterion, \eg\ on the value of the error function, because this
latter choice would
tend to artificially damp
statistical fluctuations of the pseudo--data sample, and it could thus
lead to biased results. 

\section{Results}

We will present results for the proton, deuteron and nonsinglet
structure function $F_2$.  
We have decided to train a separate neural network for the nonsinglet
in view of precision applications. 
This is because
the proton and deuteron structure functions are roughly of the
same size, while their difference is typically smaller by a factor
10. Hence, in order to achieve a 1\% accuracy on  $F_2^p-F_2^d$ we
would  need approximately a 0.1\% accuracy on $F_2^p$ and $F_2^d$.
An individual fit of the proton, deuteron, and nonsinglet allows us to
achieve a comparable accuracy on all structure functions.
After discussing the general aspects of the
assessment of the quality of the fits, we will discuss separately
the nonsinglet, proton and deuteron fits.

\subsection{Fit assessment}

As discussed in Sect.~4.1, the determination of the probability
density in the space of structure function is based on
the training of  a neural network on each
of the replicas of the original data set. We must therefore first,
assess whether the training of each network is sufficient, and then
verify that the set of networks represents the probability density of
the data in a faithful and unbiased way.

The goodness of fit provided by each net is measured by
the error function $E^{(k)}$ 
Eq.~(\ref{diaenergy}) which is minimized by the training process; the
average goodness of fit is thus
\beq
\langle
E \rangle_{rep}=\frac{1}{N_{rep}}\sum_{k=1}^{N_{rep}}
\frac{1}{N_{dat}}  E^{(k)}
\label{avendef}
\eeq
where we have further normalized to the number of data so that
Eq.~(\ref{avendef}) coincides with the mean square error per data
point per replica. We also define  the normalized error function computed for
the original set of data (rather than one of the replicas) 
\beq
E^{(0)}=\frac{1}{N_{dat}}
\sum_{i=1}^{N_{dat}} \frac{\l( F^{(exp)}_i-F^{(net)(0)}_i \r)^2}
{{\sigma^{(exp)}_{i,s}}^2},
\label{zeroendef}
\eeq
where $F^{(net)(0)}_i$ is the prediction from a neural net trained on
the set of central experimental values $F^{(exp)}_i$.

The estimator for the central value of each data point is
given by the expectation value $\blan F_i^{(net)}\bran_{rep}$
Eq.~(\ref{valesdef}). The goodness of the fit to the original data
provided by this estimator is  
\beq
\chi^{2} = \frac{1}{N_{dat}}\sum_{i,\,j=1}^{N_{dat}} 
\left(\Big\lan F_i^{(net)}\Big\ran_{rep} -  
F_i^{(exp)}\right) \mathrm{cov}^{-1}_{ij}
\left(\Big\lan F_j^{(net)}\Big\ran_{rep} -  
F_j^{(exp)}\right),
\label{chisqdef}
\eeq
where $\mathrm{cov}_{ij}$ is the covariance matrix Eq.~({\ref{covdef}), and we
have again normalized to the number of data points. Note that 
the number of degrees of freedom is not just the difference of
$N_{dat}$
and the number
of parameters of the neural network, because  the neural
network is by construction redundant (see Sect.~4.3): 
a network with a smaller number of parameters could lead to 
fits of the same quality, though with a
longer training. However,  the data points are about 500 
(see Sect.~2) while
the parameters of the net are about 50 (see Sect.~4.3), corresponding
to some 10 or 20 truly independent degrees of freedom: hence, the expression
Eq.~(\ref{chisqdef}) 
differs only by a few percent from the $\chi^2$ per degree of freedom.

In order to assess how well experimental errors are
reproduced by their neural estimator
Eq.~(\ref{sigesdef}) we define the corresponding
average variance and percentage errors
\bea
\Big\lan V\l[\Big\lan F^{(net)} \Big\ran_{\ngen}\r]\Big\ran_{dat} &=& 
\frac{1}{N_{dat}} \sum_{i=1}^{N_{dat}}\l(\Big\lan F^{(net)}_i\Big\ran_{\ngen}-F^{(exp)}_i\r)^2 \\
\Big\lan PE\l[\Big\lan F^{(net)} \Big\ran_{\ngen}\r]\Big\ran_{dat}
&=& 
\frac{1}{N_{dat}} \sum_{i=1}^{N_{dat}}\l|\frac{\Big\lan F^{(net)}_i\Big\ran_{\ngen}-F^{(exp)}_i}{F^{(exp)}_i}\r|\,. 
\eea
We can similarly define percentage errors on the 
correlation and covariance 
$\Big\lan V[\rho^{(net)}]\Big\ran_{dat}$
and $\Big\lan V[\mathrm{cov}^{(net)}]\Big\ran_{dat}$. 

A point--by--point measure of the agreement between network estimators
and their experimental counterparts is provided by defining  the
scatter correlation of central values, in analogy to Eq.~(\ref{sccdef})
\beq
r[F^{(net)}]=\frac{\Big\lan F^{(exp)}
 \Big\lan F^{(net)}\Big\ran_{\ngen}  
\Big\ran_{dat} - \Big\lan F^{(exp)}\Big\ran_{dat}
\Big\lan \Big\lan F^{(net)}\Big\ran_{\ngen}  
\Big\ran_{dat}}{\sigma_s^{(exp)}\,\sigma_s^{(net)}}\,.
\label{sccndef}
\eeq
where the scatter variance of the network is
\beq
\sigma_s^{(net)}= \blan \left(\blan F^{(net)}\bran_{rep}\right)^2\bran_{dat}-
\left(\blan\blan F^{(net)}\bran_{rep}\bran_{dat}\right)^2.
\label{scsndef}
\eeq
Similarly we define $r[\sigma^{(net)}]$, 
$r[\rho^{(net)}]$ and
$r[\mathrm{cov}^{(net)}]$. As discussed in Sect.~4.2, 
the scatter correlation $-1\le r\le1$ indicates how
closely the network estimator follows its experimental counterpart as
the index $i$ runs over experimental points. Specifically, $r[\sigma^{(net)}]\approx1$
implies that $\langle \sigma^{(net)}_i\rangle$  is approximately 
proportional to
$\sigma^{(exp)}_i$
(and similarly for  $r[F^{(net)}]$ etc.). The coefficient of
proportionality can then be found by comparing the average
experimental variance Eq.~(\ref{expave}) with the average network variance
\beq
\blan \sigma^{(net)}\bran_{dat}=\frac{1}{N_{dat}}\sum_{i=1}^{N_{dat}}
\sigma^{(net)}_i. 
\label{avvardefn}
\eeq
A substantial deviation
of the scatter correlation
from one indicates that the neural estimator is not following
the pattern of experimental variations. This might indicate that
the neural network is incapable of reproducing the shape of the data,
or else that it is successfully reducing fluctuations in the data.

In practice, we are interested in the case in which the neural
networks lead to a successful fit (\ie~with a good value of the
$\chi^2$~(\ref{chisqdef})), while leading to a reduction of the
variance, as discussed in Sect.~4.1, so that $\langle
\sigma^{(net)}\rangle_{dat}< \langle\sigma^{(exp)}\rangle_{dat}$. 
Because we are minimizing the diagonal
error function Eq.~(\ref{diaenergy}) and not the full likelihood, 
we cannot simply rely on the value of the average error $\langle
E\rangle$ and the $\chi^2$ to assess the quality of the fit. Rather,
we must verify that the distribution of
systematic errors remains unbiased, and that if a reduction of the
variance is observed it is not due to  lack of flexibility of the
neural network.

In order to test whether systematics are reproduced in an unbiased
way, 
we notice that experimental correlations are entirely
due to systematic errors, according to
Eqs.~(\ref{cornmc}-\ref{corbcd}). 
It follows from these equations and from the definition of covariance
matrix~(\ref{covdef}) that if the networks reduce the statistical
variance but reproduce the experimental systematics (as they ought
to), then $\sigma_{i}^{(net)}< \sigma_{i}^{(exp)}$ and
$\rho_{ij}^{(net)}> \rho_{ij}^{(exp)}$ but 
$\mathrm{cov}_{ij}^{(net)}\approx \mathrm{cov}_{ij}^{(exp)}$. So if
$\mathrm{cov}^{(net)}$ and $\mathrm{cov}^{(exp)}$ are
strongly correlated and have approximately the same mean, the
systematics are reproduced in an unbiased way. Notice
however that this condition is in general sufficient, but not
necessary: for instance, more measurements at the same point
from independent experiments have $\rho^{(exp)}=0$ identically, while
$\rho^{(net)}=1$, so in this case $\mathrm{cov}^{(exp)}$ and 
$\mathrm{cov}^{(net)}$ cannot possibly be equal.

In order to test whether the variance reduction is due to the fact
that the networks are distributed with smaller variance around the
true experimental values, we construct  a suitable estimator, assuming
for simplicity all errors to be Gaussian.
Consider a measurement value $m_i$ of $F_2$, where $i$
represents a pair of values $(x, Q^2)$. For each measurement
\beq
m_i=t_i+\sigma_i\,s_i\,
\label{flucm}
\eeq
where $s_i$ is a univariate Gaussian number with zero mean, 
$t_i$ is the true value of $F_2$, and $\sigma_i$ its error.
The $k^{th}$ replica of generated data is then
\beq
g_i^{(k)}=m_i+r^{(k)}_i\,\sigma_i=t_i+(s_i+r^{(k)}_i)\sigma_i\,,
\label{flucg}
\eeq
where $r^k_i$ is also a univariate zero mean Gaussian random number. 
Assume now that the best--fit neural networks are distributed about 
the true values $t_i$
with an error $\hat\sigma_i$. For
the $k^{th}$ neural network we have
\beq
n_i^{(k)}=t_i+{r'}^{(k)}_i\,\hat\sigma_i\,,
\label{flucn}
\eeq
where, assuming that the network is an unbiased estimator,
${r'}^{(k)}_i$ have zero mean, and can be normalized to be univariate. Because
the networks are determined by fitting to $g_i^{(k)}$, in
general, ${r'}^{(k)}_i$ will be correlated to $r^{(k)}_i$ and $t_i$.

If we assume that the number of replicas is large enough that
$\blan r^{(k)}_i\bran_{rep}=\blan {r'}^{(k)}_i\bran_{rep}=0$ 
and $\blan (r^{(k)}_i)^2\bran_{rep}=\blan
({r'}^{(k)}_i)^2\bran_{rep}=1$, while the number of data points is
large enough that $\langle s_i\rangle_{dat}=0$
  and $\langle s_i^2\rangle_{dat}=1$ 
we get the average error Eq.~(\ref{avendef})
\beq
\langle E \rangle_{rep}= 2+ \langle
(\hat\sigma/\sigma)^2\rangle_{dat}
- 2 \langle \langle r r'\rangle_{rep} (\hat\sigma/\sigma)\rangle_{dat},
\label{avenres}
\eeq
where  $\langle\rangle_{dat}$ denotes averaging with
respect to data in one replica
(index $i$) and $\langle\rangle_{rep}$ denotes averaging with
respect to replicas (index $k$).
Now, define a modified average error, where the prediction of
the $k$--th network instead of being compared to the $k$--th replica (as
in the average error proper Eq.~(\ref{avendef}))
are compared to the experimental data:
\bea
\widetilde{E}^{(k)}&=&
\sum_{i=1}^{N_{dat}} \frac{\l( F^{(exp)}_i-F^{(net)(k)}_i \r)^2}
{{\sigma^{(exp)}_{i,s}}^2};\\
\langle 
\widetilde{E} \rangle_{rep}&=&\frac{1}{N_{rep}}\sum_{k=1}^{N_{rep}}
\frac{1}{N_{dat}}  \widetilde{E}^{(k)}
\label{modavendef}
\eea
We get immediately
\beq
\langle \widetilde{E} \rangle_{rep}= 1+ \langle
(\hat\sigma/\sigma)^2\rangle_{dat}.
\label{modavenres}
\eeq

We finally define the ratio
\beq
{\cal R}\equiv \frac{\langle \widetilde{E}\rangle_{rep} }{\langle  E\rangle_{rep}},
\label{calrdef}
\eeq
which is the desired estimator. Indeed, assume that the networks
displays significantly smaller fluctuations than the data,
$\sigma_i^{(net)}<\!<\sigma_i^{(exp)}$. We  wish to test whether this
is due to the fact that the network has
performed  substantial error reduction by combining several experimental
data points $g_i^{(k)}$ into a determination of its parameters. 
In such case the correlation between
${r'}_i^{(k)}$ and $r_i^{(k)}$  is very weak, since ${r'}_i^{(k)}$ 
is determined by the
values of several points $g_j^{(k)}$ with $j\not= i$. Hence
\beq
{\cal R}\approx \frac
{1+\langle(\hat\sigma/\sigma)^2\rangle_{dat}}
{2+\langle(\hat\sigma/\sigma)^2\rangle_{dat}}.
\label{appcalrdef}
\eeq
If error reduction is substantial and
$\langle(\hat \sigma/\sigma^2)\rangle_{dat}<\!<1$, then
 ${\cal
R}\approx \frac{1}{2}$. Physically, this means that by combining many
data, the network manages to always be closer to the true values $t_i$
than the replicas. If instead the network was just artificially  
smoothing  the data, we would expect $(\hat \sigma_i/\sigma_i)>1$ and again no
correlation  $\blan  {r'}_i^k r_i^k\bran_{rep}\approx 0$, so ${\cal
R}\gsim 1$. 

It is instructive to consider also a case in which we observe no
error reduction, $\sigma_i^{(net)}\approx \sigma_i^{(exp)}$. In this
case, we can simply check that the neural network are behaving properly by
verifying that the scatter correlation $r[\sigma]$ is high. However,
the value of $\cal R$ provides a cross--check.  In such
case, we expect $\hat \sigma_i\approx\sigma_i$, and a large correlation
$\langle {r'}_i^k r_i^k \rangle_{rep}\approx 1$ since 
$n_i^{k}$ is determined by $g_i^{k}$. Hence, in this case we would
expect ${\cal R}\approx 2$. If instead the networks were not
reproducing the data accurately we would again get ${\cal
R}\lsim 1$ as above.

\subsection{Nonsinglet}

The nonsinglet structure function is defined as
\beq
F_2^{NS}(x,Q^2)=F_2^p(x,Q^2)-F_2^d(x,Q^2).
\label{nsdef}
\eeq
Data for this quantity are obtained by discarding the BCDMS proton
data taken with a 100~GeV beam, which have no deuteron counterpart,
and then exploiting the fact that all remaining BCDMS and NMC data
come in proton, deuteron pairs, taken at the same values of $x$ and
$Q^2$. For NMC data, there are small difference in the binning in
$Q^2$ between proton and deuteron, which are however negligible on the
scale of
the typical variation of the structure function and the size of the
experimental errors.
\begin{figure}
\begin{center}
\epsfig{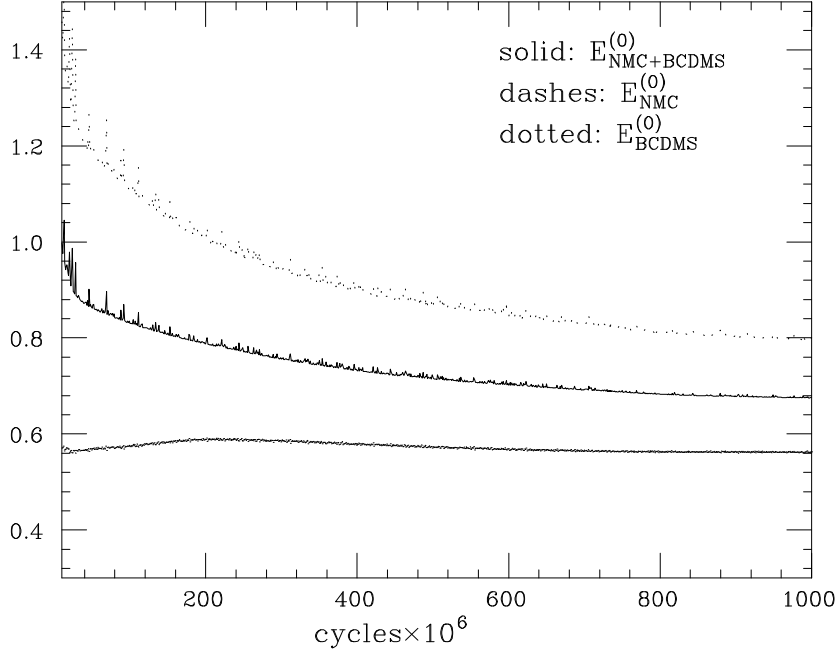} 
\end{center}
\caption{Nonsinglet: dependence  of the error on the length of training.}
\end{figure}
 
In order to establish the length of the training, we study the
behaviour of the average error function $E^{(0)}$
Eq.~(\ref{zeroendef}) 
determined
from the original set of data as a function of
the number of training cycles (Figure~6).\footnote{Cyclic
patterns in the behaviour of $E^{(0)}$ as a function of the training
length are due to the periodicity of the pseudo-random number generator
which is used to determine the permutation of data in the training
process, see Sect.~4.3.}It is seen that
an extremely long training (of order of $10^9$ training cycles) is
required in order for the error function to stabilize. However, if we
look at the average error function computed from data of either of the
two experiments, it is apparent that it is only the contribution to
$E^{(0)}$ from the BCDMS data which keeps improving, while
$E^{(0)}_{NMC}$ 
attains very rapidly a small value which is then stable. 
If we train the nets only to
data of one experiment (Figures 7--8), the training required for
reaching convergence with 
BCDMS is still rather longer than with NMC data, but  substantially
shorter than when both experiments are fitted at once. The greater
difficulty in learning the BCDMS data is due to fact that these data
have smaller errors. Notice  that the
net trained on each of the two experiments predicts to a large extent the other
experiment once convergence has been achieved. This means that,  even
though only one experiment is used for training, the contributions to $E^{(0)}$
from both experiments are reduced in the process --- though, of course,
the experiment which is not used has eventually a worse average $E^{(0)}$.
\begin{center}\begin{figure}
\begin{minipage}[t]{0.48\textwidth}
\includegraphics[width=\textwidth,clip]{chi2_nmc-ns.ps}
\caption{NS training on NMC data.}
\end{minipage}
\begin{minipage}[t]{0.48\textwidth}
\includegraphics[width=\textwidth,clip]{chi2_bcd-ns.ps}
\caption{NS training on BCDMS data.}
\end{minipage}
\end{figure}
\end{center}

\begin{figure}
\begin{center}
\epsfig{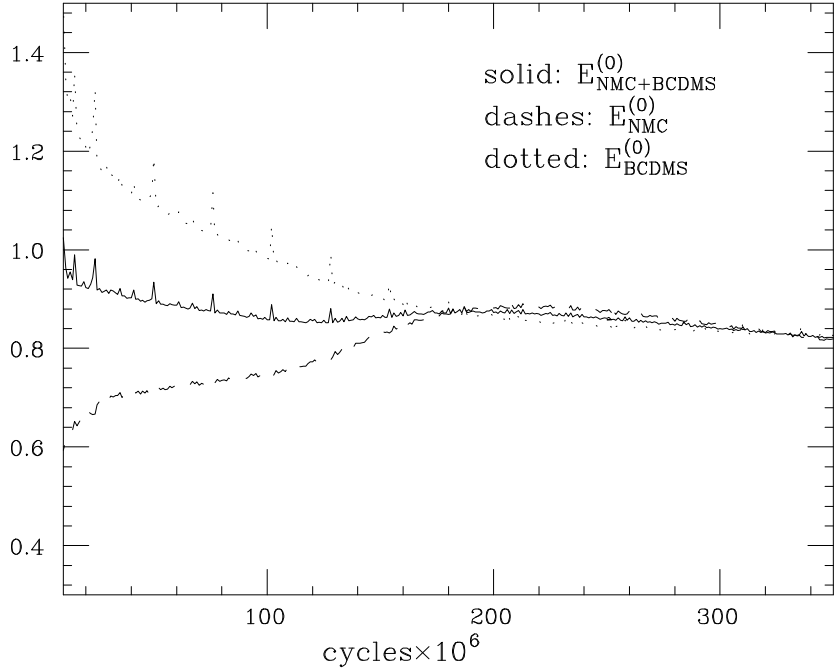} 
\end{center}
\caption{Nonsinglet weighted training (90\% BCDMS, 10\% NMC)}
\end{figure}
This suggests that an optimized training can be obtained by showing to
the network BCDMS  data more often than NMC data. Then, the BCDMS data
are learnt faster but the information from the NMC data is not
lost. An optimal combination is found (by trial and error) 
when the BCDMS data are shown
90\% of times, and NMC data 10\% of times. Then, the average error $E^{(0)}$
for the two experiments both reach convergence at the same value after
about  $1.8\times 10^8$ training cycles 
(Figure~9). Note that, after convergence has been reached, the average
error $E^{(0)}$ of the two experiment oscillate about each other,
while slowly improving at the same rate. This training is
optimized, in comparison to that of 
Figure~6 both because it is by
more than a factor~5 shorter, and, more importantly,
because the convergence of the values of
$E^{(0)}$ for the two experiments shows that the global minimum has
been found.

Indeed, a substantial
difference in  values
of $E^{(0)}$ for the two data sets is not acceptable, because it
entails that if on  average $E^{(0)}\approx 1$, in actual fact the net
is overlearning one of the two data sets, and underlearning the
other, \ie,  overlearning in one kinematic region and
underlearning in another region. 
This is what happens with $\sim 10^7$ training
cycles (origin of the plot of Fig.~6). Unless there are reasons to
believe that errors are not correctly estimated for one of the
data sets, this  means that the neural network is getting trapped
in a local minimum corresponding to the data set which has been learnt
faster, and thus unable to find the global minimum. Indeed, when the
global minimum is found, the values of $E^{(0)}$ for the two data sets
are approximately equal and 
improve at the same rate, as it happens for the weighted training of
Fig.~9 with $\gsim 2\times 10^8$ training cycles. 
It is clear from
Fig.~6 that if an unweighted
training is adopted, it might still be 
possible to achieve comparable values of 
$E^{(0)}$ for the two data sets, but only with a very long training,
leading to a value of $E^{(0)}<\!<1$. In such case, the net would be
overlearning throughout the full data region. Hence, weighted training
is mandatory if we wish to avoid both local minima and overlearning.

As discussed in Sect.~4.3, it turns out to be convenient to
let each network first undergo a preliminary short training to the simple
error function Eq.~(\ref{simpenergy}) with large
learning rate $\eta$ Eq.~(\ref{delpars}). After some experimentation
along the lines discussed above, the training parameters 
have been chosen to be as follows:   $4.6\times
10^6$ cycles with learning rate $\eta=4\times 10^{-3}$ and error function
Eq.~(\ref{simpenergy}), followed by $1.8\times 10^8$ cycles with $\eta=4\times
10^{-8}$ 90\% of the
times on BCDMS data and 10\% of the times on NMC data with error
function
Eq.~(\ref{diaenergy}). The momentum term Eq.~(\ref{momterm}) is always
set to $\alpha=0.9$.

\begin{figure}
\begin{center}
\epsfig{width=0.6\textwidth,figure=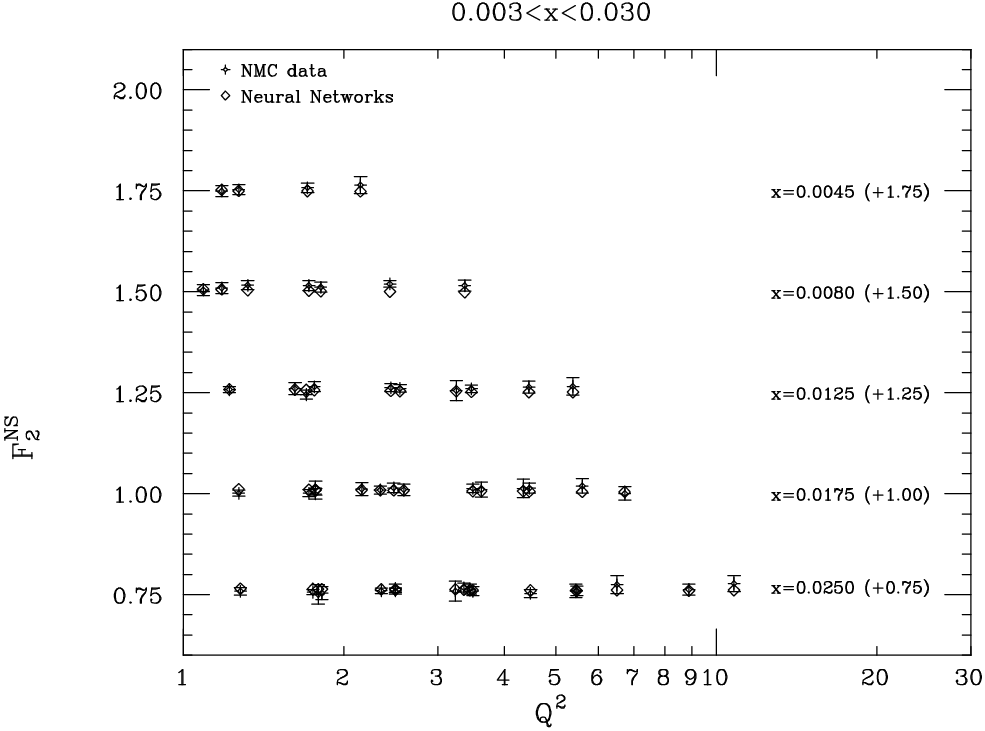} 
\epsfig{width=0.6\textwidth,figure=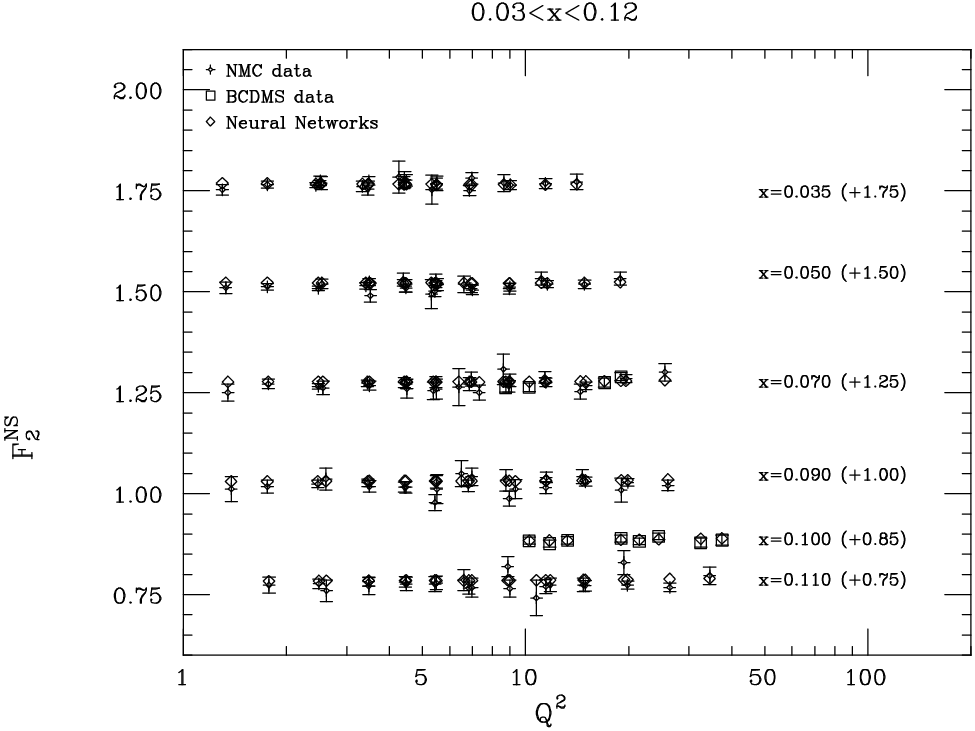} 
\epsfig{width=0.6\textwidth,figure=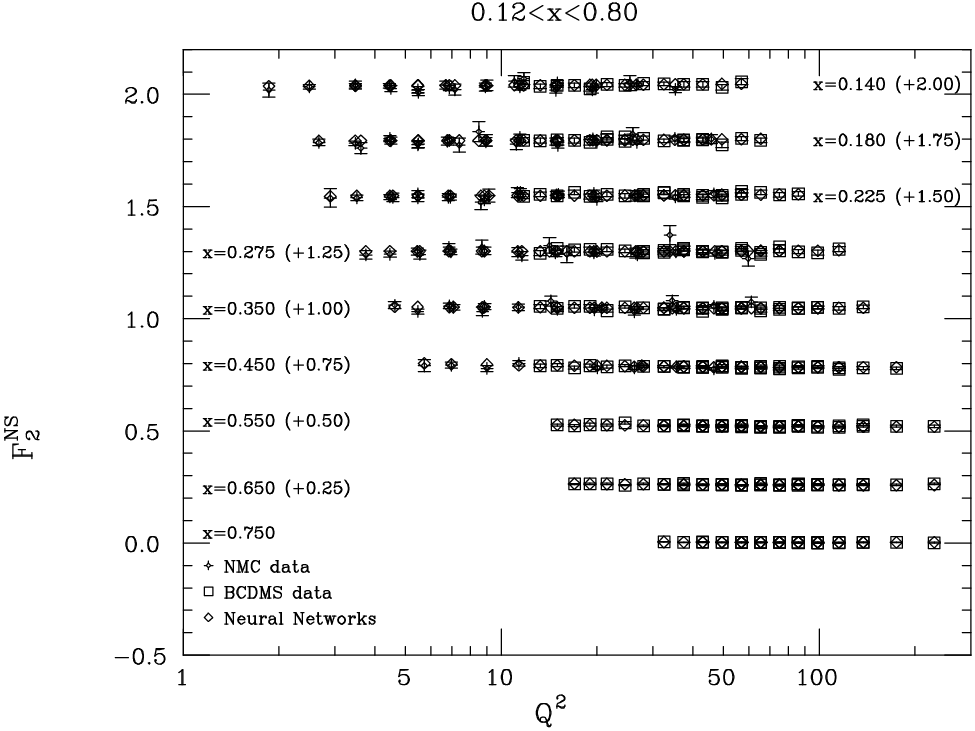} 
\end{center}
\caption{Nonsinglet data and network prediction at the corresponding $(x,Q^2)$
value.  For clarity, the points are offset by the amount given in
parenthesis.}
\end{figure}
We have thus produced a set of 1000 networks, each trained 
on a replica of the original data set. The prediction for
the structure function and its error computed from them according to
Eqs.(\ref{valesdef}-\ref{sigesdef}) are compared to the experimental 
data in 
Figure.~10.
The features of this set of neural networks are summarized in Table~4.
\begin{table}[t]  
\begin{center}  
\begin{tabular}{cccc} 
\multicolumn{4}{c}{}\\  
\hline
$N_{net}$ & NMC+BCDMS & NMC & BCDMS\\
\hline  
$\chi^{2} $ 
  & 0.79 & 0.78 &  0.80\\ 
$\lan E \ran_{rep}$ 
  & 1.14 &  1.05 &  1.25\\
${\cal R}$ 
  & 0.58 & 0.53 & 0.66\\
$r\l[F^{(net)}\r]$ 
  & 0.81 & 0.69 &  0.95\\
\hline
$\Big\lan V[\sigma^{(net)}]\Big\ran_{dat}$ 
  & $1.7\times 10^{-4}$ & $2.9\times 10^{-4}$ & $3.5\times 10^{-5}$  \\
$\Big\lan PE[\sigma^{(net)}]\Big\ran_{dat}$ 
  & 80\%  & 82\% & 77\%\\
$\Big\lan \sigma^{(exp)} \Big\ran_{dat}$ 
  & 0.011  &  0.016 & 0.006 \\
$\Big\lan \sigma^{(net)} \Big\ran_{dat}$ 
  & 0.004  & 0.005 &  0.003\\
$r[\sigma^{(net)}]$ 
  & 0.50 & -0.02 &  0.92\\
\hline
$\Big\lan V[\rho^{(net)}]\Big\ran_{dat}$ 
  & 0.34 & 0.33 &  0.35\\
$\Big\lan \rho^{(exp)} \Big\ran_{dat}$ 
  & 0.09 & 0.04 &   0.16\\
$\Big\lan \rho^{(net)} \Big\ran_{dat}$ 
  & 0.57 & 0.45 &  0.73\\
$r[\rho^{(net)}]$ 
  & 0.37 & 0.12 &  0.56\\
\hline
$\Big\lan V[\mathrm{cov}^{(net)}]\Big\ran_{dat}$ 
  & $9.7\times 10^{-10}$ & $ 1.7\times 10^{-9}$ & $ 2.1\times 10^{-11}$ \\
$\Big\lan \mathrm{cov}^{(exp)} \Big\ran_{dat}$ 
  & $8.4\times 10^{-6}$ & $ 9.7\times 10^{-6}$  & $ 6.8\times 10^{-6}$ \\
$\Big\lan \mathrm{cov}^{(net)} \Big\ran_{dat}$ 
  & $9.0\times 10^{-6}$ & $ 1.1\times 10^{-5}$  & $ 6.8\times 10^{-6}$ \\
$r[\mathrm{cov}^{(net)}]$ 
  & 0.26 & 0.21 &  0.86\\
\hline
\end{tabular}
\end{center}
\caption{Nonsinglet results}
\end{table}

The table shows that central values are very well reproduced. 
The value of $\chi^2$ of the two data sets is almost the same, and
close enough to $\chi^2\approx1$, so that there is neither
overlearning nor underlearning.
The
statistical error is substantially reduced by the network, especially
for NMC data which have larger experimental error. Correlations
correspondingly increase, so that the average covariance is
essentially unchanged: this indicates that the systematics is
reproduced in an unbiased way. The expectation that ${\cal R}\approx
0.5$ is beautifully fulfilled, indicating that the variance reduction
is due to the fact that the network is learning an underlying law. 

It is interesting to observe  
that while $\sigma^{(net)}$ is highly correlated to
$\sigma^{(exp)}$ for
BCDMS data, this does not happen for  NMC. The reason for this can be
traced to the fact that 
local information provided by each NMC data point is very weak:
a few NMC points on top of the BCDMS data  are sufficient to
train the neural network and the remaining ones do not provide
significant extra information.  To illustrate this, in 
Figure~11 we
show the average error computed from a network trained on
all BCDMS data, but only 20 NMC
points (7\% of NMC points arbitrarily chosen among those where the
systematics is less than the statistical error):
this fit is as good as the fit where all NMC data are kept. Note,
however, that in this case the value of $E^{(0)}$  computed from the
two data sets cross and rapidly diverge as the number of cycles grows
above $1.5\times 10^8$. This divergence is not observed when the full
set of NMC data is included (Figure~9): this shows that
the reduced NMC data set does not carry the full information needed for
a detailed fit, so the
use of the full data set remains necessary if we want to avoid fine--tuning.
\vfill\eject

It is clear from Fig.~10 that the coverage of the data is such that,
within the data region, the interpolating structure function is very
well constrained by the data. In fact, the observed substantial error
reduction is due to the fact that each new data point modifies very
little the behaviour predicted by neighbouring data. It is therefore
interesting to ask how far one can trust the neural networks when
extrapolating outside the data region. 

At large $x$, it can be easily
verified that the extrapolation is sufficiently
constrained by the kinematic bound $F_2(1,Q^2)=0$ that no  increase in
the uncertainty is observed when $x$ goes outside the region of the
data even at $Q^2=3$~GeV$^2$, where there are no data with $x>0.3$.
In the small $x$ region, instead, the uncertainty increases very
rapidly: for instance, at $Q^2=30$~GeV$^2$ the error is $\sigma\approx
0.004$ at $x=0.3$ (bulk of the data), $\sigma\approx
0.006$ at $x=0.1$ (edge of the data region) and
$\sigma\approx 0.017$ at $x=0.01$ (extrapolation by one order of
magnitude in $1/x$). Similar behaviour is observed at other scales:
hence, extrapolation at small $x$ outside the data region is not
predictive. 

Coming now to extrapolation in $Q^2$, it turns out that
extrapolation at large $Q^2$ is somewhat more reliable than
extrapolation at low $Q^2$. For instance, at $x=0.3$ the error is 
$\sigma\approx
0.004$ at $Q^2=30$~GeV$^2$ (bulk of the data). At low $Q^2$ it reaches 
$\sigma\approx
0.006$ at $Q^2=3$~GeV$^2$ (edge of the data), and becomes $\sigma\approx
0.016$ at $Q^2=1$~GeV$^2$ (extrapolation). At high $Q^2$ it reaches 
$\sigma\approx
0.004$ at $Q^2=100$~GeV$^2$ (edge of the data), and becomes $\sigma\approx
0.007$ at $Q^2=300$~GeV$^2$ (extrapolation). Similar behaviour is
observed for other values of $x$. The fact that the extrapolation is
more stable at large $Q^2$ can be understood as a consequence of the
fact that, because of
asymptotic freedom,  the large $Q^2$ scaling violations are smaller
and smoother, and thus easier for the neural network to capture. Note however
that, in contrast to a parton fit,
here  the theoretical information on scaling
violations encoded in evolution equations is not being used, and thus 
extrapolation, even at large $Q^2$, should be used with care.

\begin{figure}
\begin{center}
\epsfig{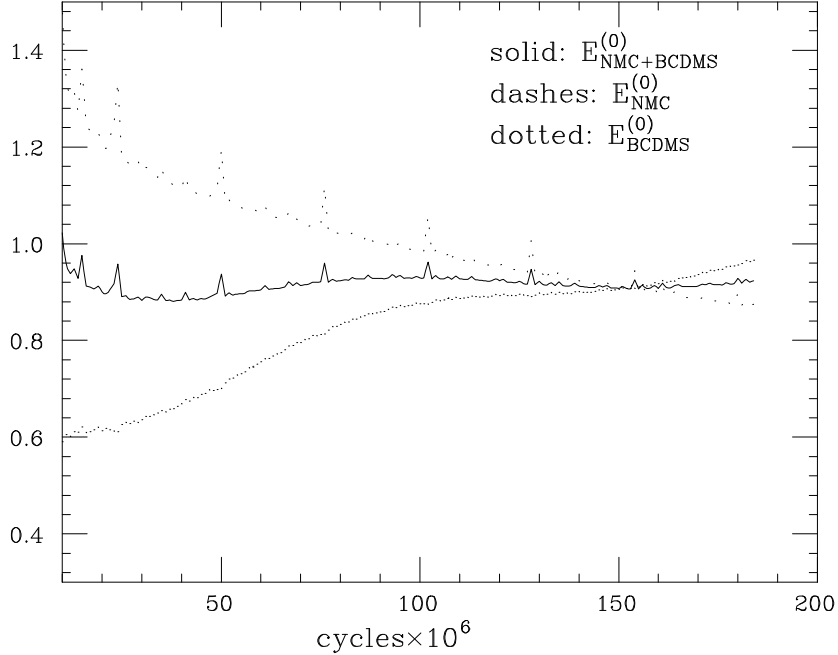} 
\end{center}
\caption{Nonsinglet training on all  BCDMS data and 20 NMC data}
\end{figure}

\subsection{Proton and deuteron}

Neural networks for proton and deuteron are trained independently, but
on a simultaneous set of pseudo--data. Namely, because proton and
deuteron experimental data are correlated (recall Sect.~2.1),
each pseudo--data replica is generated as a set of correlated proton
and deuteron data points. Two independent proton and deuteron 
neural networks are then trained on the proton and deuteron data of
each replica. This guarantees that experimental correlations between
proton and deuteron will be reproduced by the network sample. 

\subsubsection{Proton}
\begin{figure}
\begin{center}
\epsfig{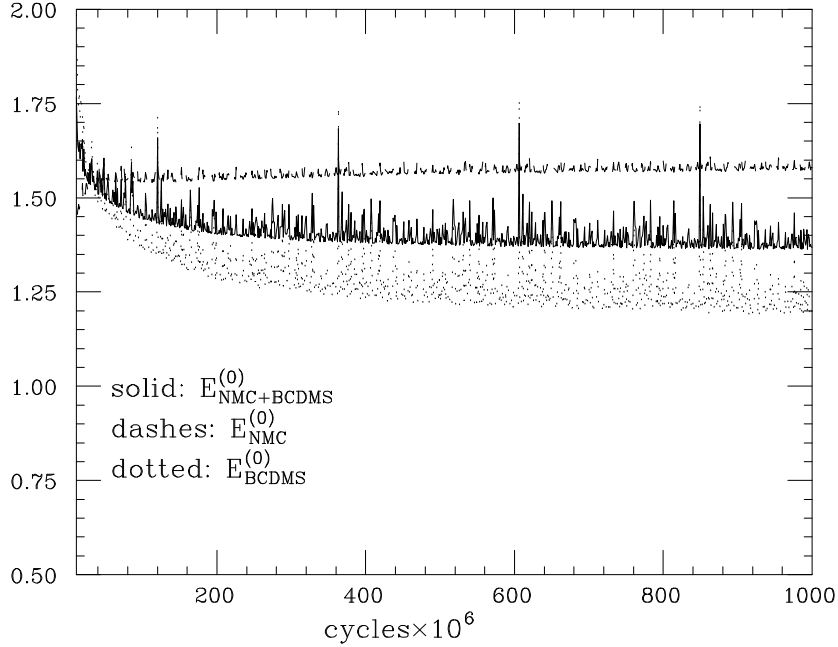} 
\end{center}
\caption{Proton: dependence  of the error on the length of training.}
\end{figure}
The training of neural networks on
proton structure function data is rather more difficult than in the
nonsinglet case, because these
data have a much smaller percentage error. The dependence of the
average error on the training length is displayed in 
Figure~12. The
noisiness of the training process reflects the high precision of the
data, especially for BCDMS. Clearly, there are two main difficulties:
first, the BCDMS data are only learnt very slowly, and second, the
quality of the fit to NMC data is poor, and does not improve (in
fact, it slowly deteriorates) as the BCDMS data are learnt.

Inspection of the fits to data of a single experiment (Figures~13--14)
reveals that, similarly to the nonsinglet case, NMC data are learnt
very fast whereas BCDMS data require a much longer training. However,
unlike in the nonsinglet case, networks trained on data
of one experiment are now entirely incapable of predicting data from the other
experiment: as one  average error slowly improves, the other deteriorates
dramatically. This suggests that an optimized training can again be
obtained by giving more weight to the BCDMS data, however the
imbalance cannot be too large because we cannot rely on one experiment
being able to predict the other. 
\begin{center}\begin{figure}
\begin{minipage}[t]{0.48\textwidth}
\includegraphics[width=\textwidth,clip]{chi2_nmc-p.ps}
\caption{Proton training with NMC data.}
\end{minipage}
\begin{minipage}[t]{0.48\textwidth}
\includegraphics[width=\textwidth,clip]{chi2_bcd-p.ps}
\caption{Proton training with BCDMS data.}
\end{minipage}
\end{figure}
\end{center}

\begin{figure}
\begin{center}
\epsfig{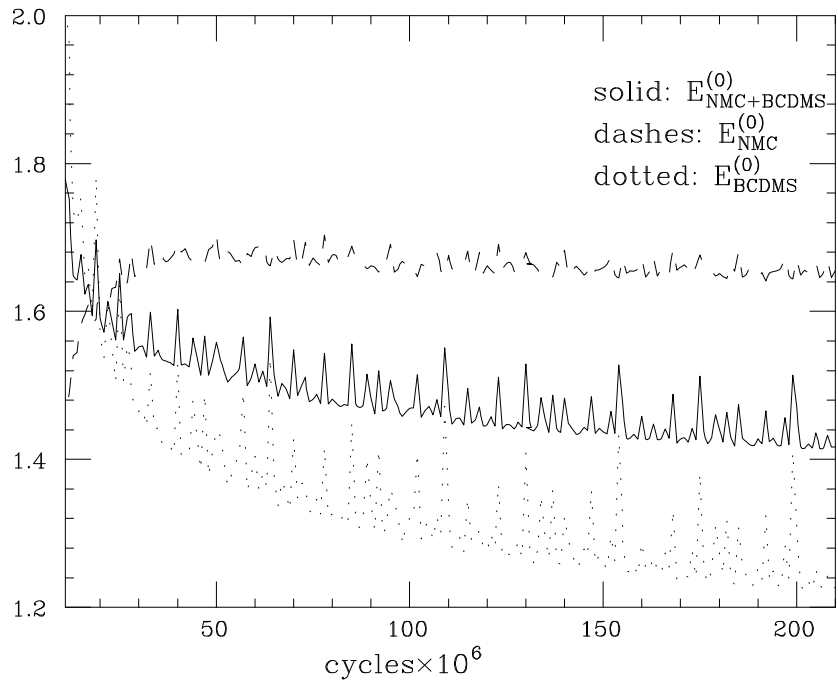} 
\end{center}
\caption{Proton weighted training (60\% BCDMS, 40\% NMC)}
\end{figure}
An optimal combination is found  when
 BCDMS data are shown 60\% of the time and  NMC data 40\% of time 
(Figure 15). Both experiments then
reach convergence after about  $2\times 10^8$ training cycles.
However, unlike in the nonsinglet case, the average error for NMC
remains significantly worse than BCDMS. In fact, 
this feature is already present in fits to a single experiment of
Figures~13--14: the  average
error for NMC at convergence, though learnt fast, remains rather worse 
than BCDMS. We found no way of obtaining better fits of NMC data even
with NMC data only, much longer training, or larger architecture of
the network. This appears therefore to be a feature of the data, to
which we will come back shortly.

It is interesting to observe that the constraint  $F_2(1,Q^2)=0$ is
crucial in order to ensure proper convergence of these fits. On the
other hand, it should be noticed that, because this constraint is only
imposed with an absolute error of $10^{-3}$, the structure function
fits in the vicinity of $x=1$ are only reliable within this accuracy,
especially at low $Q^2$ where no large $x$ data are available.
\begin{figure}
\begin{center}
\epsfig{width=0.6\textwidth,figure=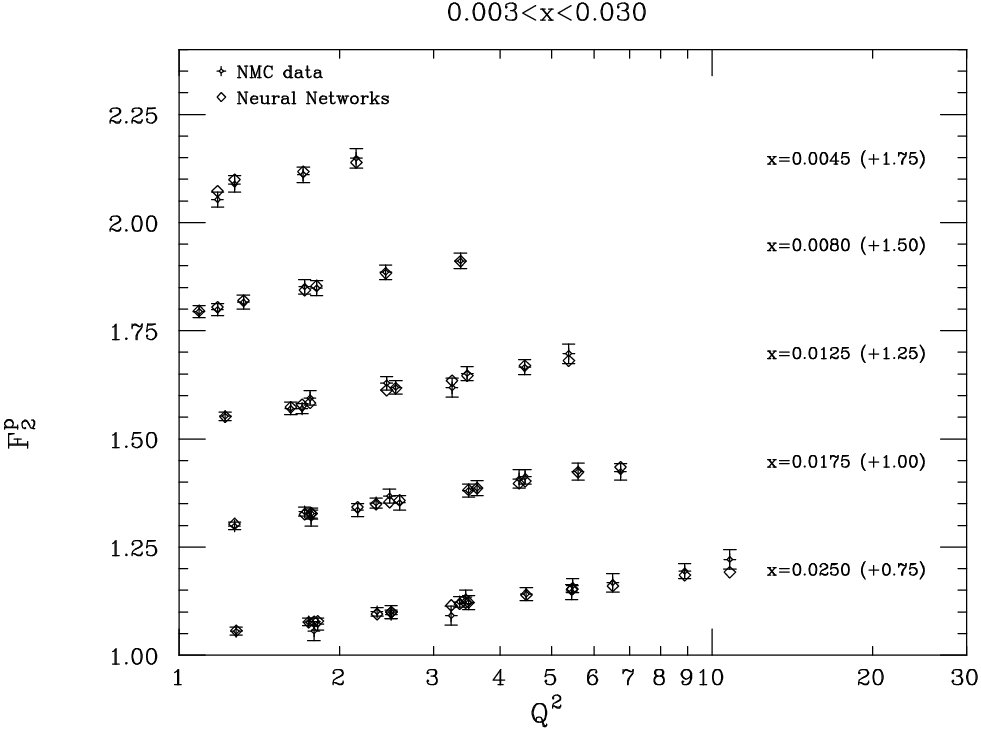} 
\epsfig{width=0.6\textwidth,figure=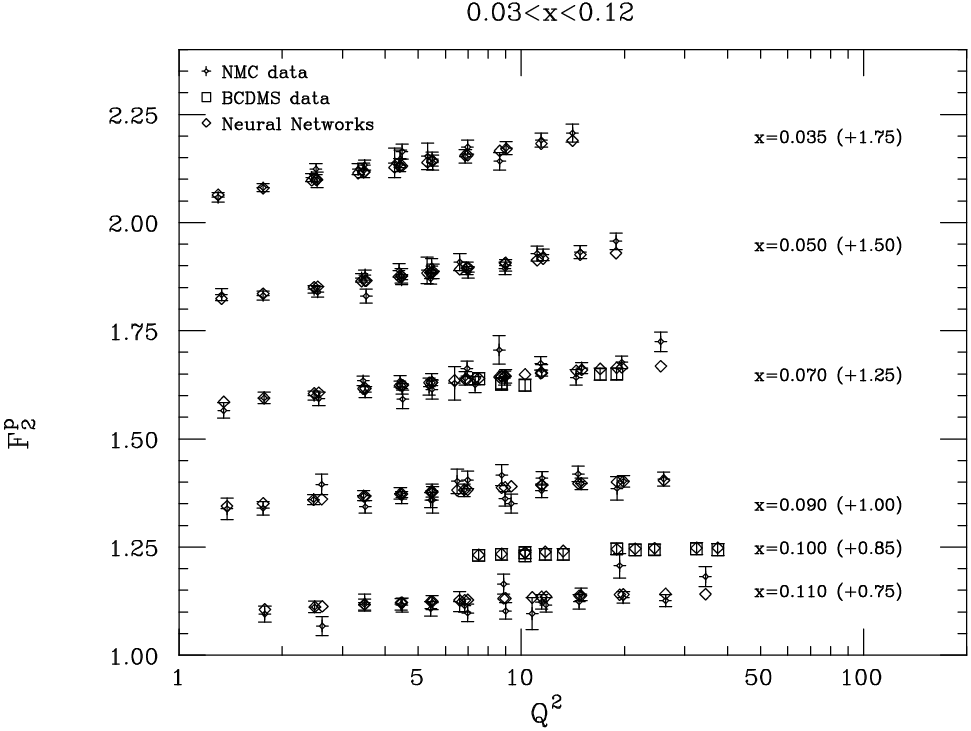} 
\epsfig{width=0.6\textwidth,figure=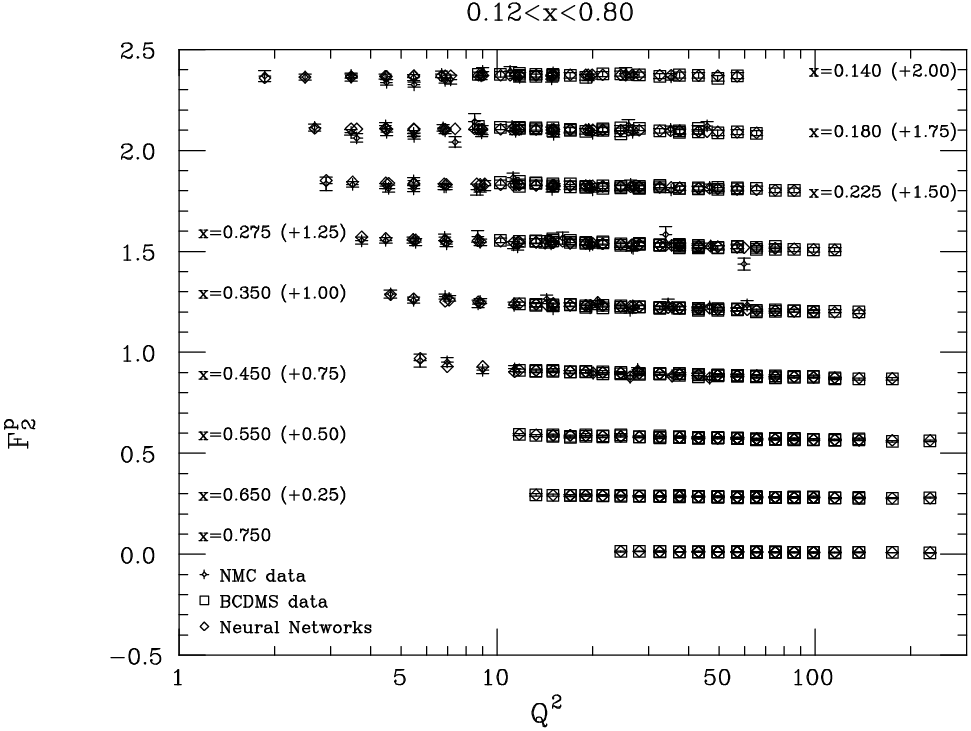} 
\end{center}
\caption{Proton data and network prediction at the corresponding $(x,Q^2)$
value.}  
\end{figure}
The final training parameters are chosen 
as follows:   $9.6\times
10^6$ cycles with learning rate $\eta=9\times 10^{-3}$ and error function
Eq.~(\ref{simpenergy}), followed by  $2\times 10^8$ cycles with $\eta=4\times
10^{-8}$ 60\% of the
times on BCDMS data and 40\% of the times on NMC data and error
function
Eq.~(\ref{diaenergy}). The momentum term Eq.~(\ref{momterm}) is always
set at $\alpha=0.9$. A sample of 1000 networks trained in this way is
used to produce values of the structure function and error which are
compared to the data in 
Figure~16. The features of these networks are
summarized in Table~5.

\begin{table}[t]  
\begin{center}  
\begin{tabular}{cccc} 
\multicolumn{4}{c}{}\\  
\hline
$N_{net}$ & NMC+BCDMS & NMC & BCDMS\\
\hline  
$\chi^{2}$ 
  & 1.30 & 1.53 & 1.12 \\ 
$\lan E \ran_{rep}$ 
  & 3.55 & 2.69 &  4.23\\
${\cal R}$ 
  & 1.45 & 0.75 & 2.00\\
$r\l[F^{net}\r]$ 
  & 0.996 & 0.956 &  0.999\\
\hline
$\Big\lan V[\sigma^{(net)}]\Big\ran_{dat}$ 
  & $1.2\times 10^{-4}$ & $2.3\times 10^{-4}$ & $2.5\times 10^{-5}$  \\
$\Big\lan PE[\sigma^{(net)}]\Big\ran_{dat}$ 
  & 50\%  & 66\% & 37\%\\
$\Big\lan \sigma^{(exp)} \Big\ran_{dat}$ 
  & 0.012 & 0.017 &  0.007\\
$\Big\lan \sigma^{(net)} \Big\ran_{dat}$ 
  & 0.007 & 0.009 &  0.006\\
$r[\sigma^{(net)}]$ 
  & 0.68 & 0.17 &  0.98\\
\hline
$\Big\lan V[\rho^{(net)}]\Big\ran_{dat}$ 
  & 0.17 & 0.27 &  0.09\\
$\Big\lan \rho^{(exp)} \Big\ran_{dat}$ 
  & 0.38 & 0.17 &  0.52\\
$\Big\lan \rho^{(net)} \Big\ran_{dat}$ 
  & 0.69 & 0.53 &  0.80\\
$r[\rho^{(net)}]$ 
  & 0.58 & 0.15 &  0.80\\
\hline
$\Big\lan V[\mathrm{cov}^{(net)}]\Big\ran_{dat}$ 
  & $1.2\times 10^{-9}$ & $2.7\times 10^{-9}$ & $ 1.3\times 10^{-10}$ \\
$\Big\lan \mathrm{cov}^{(exp)} \Big\ran_{dat}$ 
  & $3.8\times 10^{-5}$ & $4.5\times 10^{-5}$  & $3.4\times 10^{-5}$ \\
$\Big\lan \mathrm{cov}^{(net)} \Big\ran_{dat}$ 
  & $3.2\times 10^{-5}$ & $3.7\times 10^{-5}$  & $2.9\times 10^{-5}$ \\
$r[\mathrm{cov}^{(net)}]$ 
  & 0.66 & 0.36 & 0.97\\
\hline
\end{tabular}
\end{center}
\caption{Proton results}
\end{table}
\vfill\eject
Table~5 shows that all central values are well reproduced, even
though the $\chi^2$ for NMC is rather high. Errors and correlations
are very well reproduced for the BCDMS data. There does not appear to
be any reduction of variance for this experiment: $\sigma^{(net)}\approx
\sigma^{(exp)}$ and $\rho^{(net)}\approx
\rho^{(exp)}$: the networks simply reproduce the behaviour of the
data. The prediction for the ratio Eq.~(\ref{calrdef}) ${\cal R}\approx
2$  is again beautifully borne out by the data. The NMC data instead
display significant variance reduction. However, the covariance is not
reduced, indicating that systematics are faithfully reproduced, and
the value of $\cal R$ is in
good agreement with the expectation of Eq.~(\ref{appcalrdef}), 
with
$\hat\sigma=\langle \sigma^{(net)}\rangle$. The weak
scatter correlation of variances can be understood analogously to the
nonsinglet case, as a consequence of the fact that the nets can
predict most NMC data based on the information provided by a subset of
these data.

\begin{center}\begin{figure}
\begin{minipage}[t]{0.48\textwidth}
\includegraphics[width=\textwidth,clip]{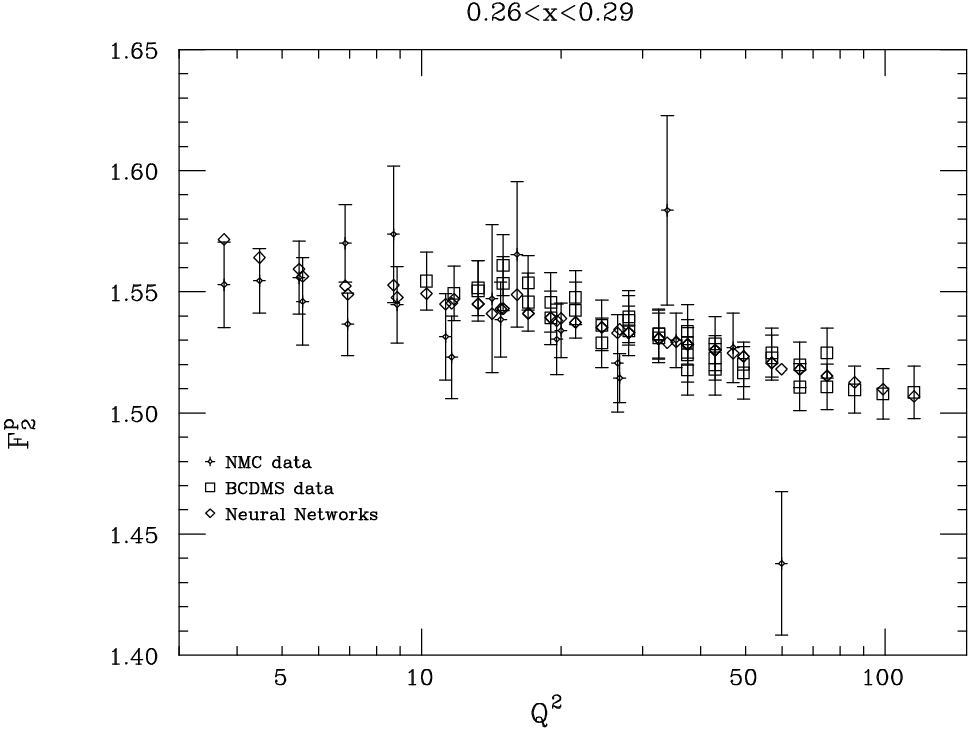}
\caption{Blow--up of proton data and predictions.}
\end{minipage}
\begin{minipage}[t]{0.48\textwidth}
\includegraphics[width=\textwidth,clip]{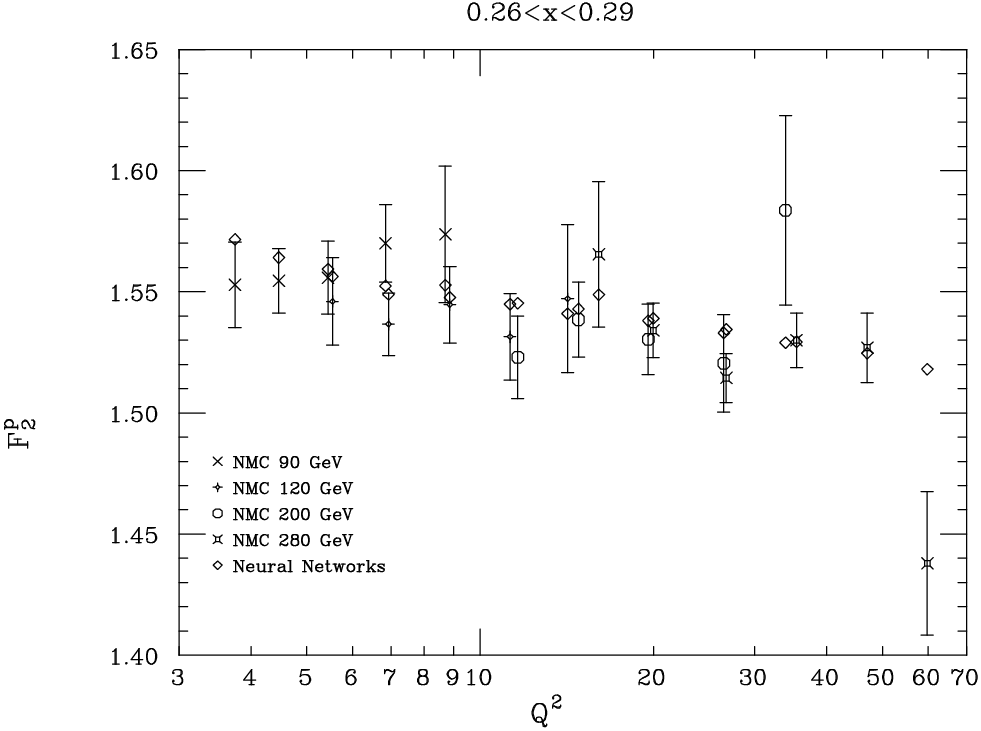}
\caption{Proton NMC data  at different beam energies, and associate 
predictions.}
\end{minipage}
\end{figure}
\end{center}
We are therefore left with the problem of the large value of $\chi^2$
for NMC. The difficulty in obtaining a satisfactory fit of NMC data
was already noted by other groups~\cite{cteq6,GKK} in the context of
global parton fits. The origin of the problem can be understood by
taking a closer look
at the NMC data, especially in the central kinematic region (middle plot
in Figure~16). It is clear that  a sizable fraction of these data are
not compatible with the remaining ones: some data points deviate by
several standard deviation from other measurements performed at equal or almost
equal values of $(x,Q^2)$. This may suggested underestimated
systematics. A blow--up of the data, however  (Figure 17), reveals
that NMC data not only disagree with BCDMS data with the same
$(x,Q^2)$, but in fact also with other NMC data at neighbouring
$(x,Q^2)$. Furthermore (Figure 18) the inconsistency persists even
within sets of data taken by NMC at fixed beam energy. Therefore, none
of the known sources of systematics can cause this
discrepancy. Inspection reveals that the data which appear to be
display the largest disagreement 
are always at the edge of a bin (\eg, the largest or
smallest $Q^2$ value for given $x$). The cumulative effect of these
inconsistencies leads to a very large $\chi^2$ which cannot be
reasonably attributed to a statistical fluctuation.

Whereas we are unable to pinpoint
the precise origin of the effect, we conclude that a  subset of the
NMC data are not consistent with remaining information on the proton
structure function, and that this inconsistency cannot be cured by
enlarging any of the experimental systematics. The neural network
correctly discards these points, and the relatively large value of $\chi^2$ 
signals this inconsistency in the data.

A consequence of the relatively poor quality of the fit to NMC proton data
is that in this case extrapolation at large $Q^2$ is as unreliable as
extrapolation at low $Q^2$, because the large $Q^2$ region is
dominated by the NMC data. Interpolation, as well as extrapolation at
large $x$ in the region where low $x$ data are available remain quite
accurate. 
\subsubsection{Deuteron}

\begin{figure}
\begin{center}
\epsfig{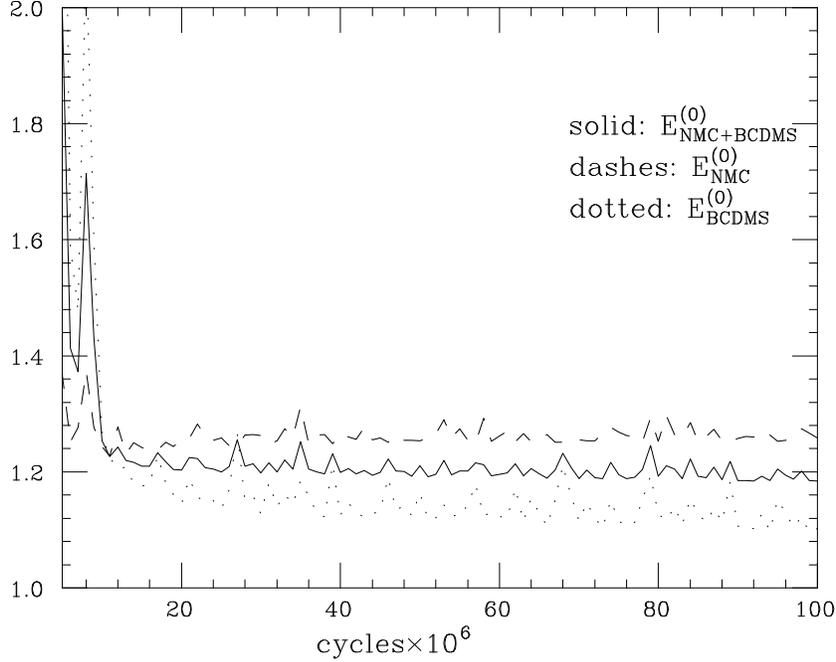} 
\end{center}
\caption{Deuteron: dependence  of the error on the length of training.}
\end{figure}
The training of networks on deuteron data does not present specific
problems. The dependence of the average error on the training length
is displayed in 
Figure~19, and shows reasonable fast convergence after
about $10^8$ training cycles. Fits to individual data sets are
displayed in Figures~20--21: BCDMS data are learnt more slowly and
have a slightly better $\langle E^{(0)}\rangle$ than NMC data. Training on
individual data sets shows that the two data sets cannot be used to
predict each other, in that the average error of the predicted set is
poor, even though it does not deteriorate as the training
progresses. No weighted training is necessary.
\begin{center}\begin{figure}
\begin{minipage}[t]{0.48\textwidth}
\includegraphics[width=\textwidth,clip]{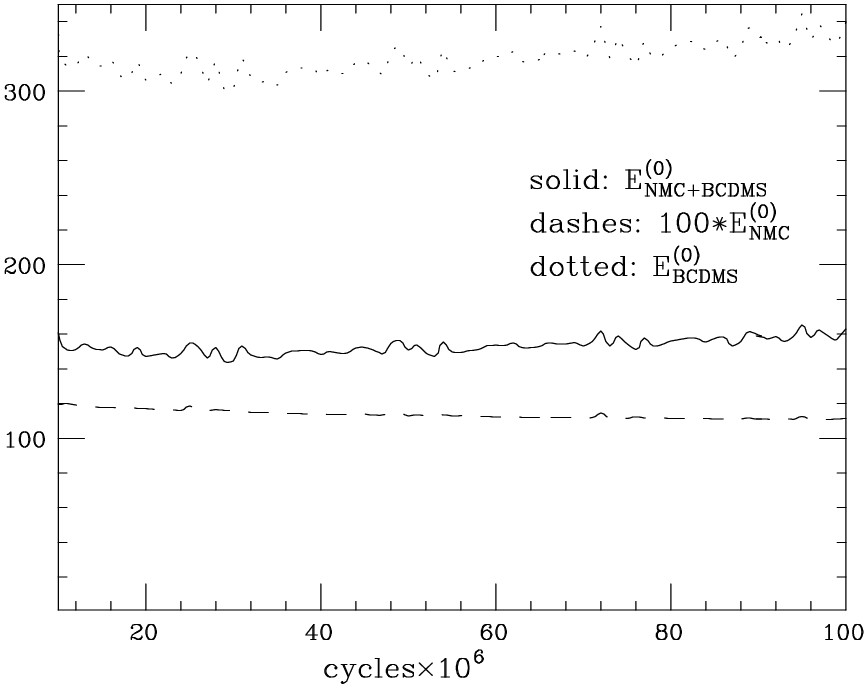}
\caption{Deuteron training with NMC data.}
\end{minipage}
\begin{minipage}[t]{0.48\textwidth}
\includegraphics[width=\textwidth,clip]{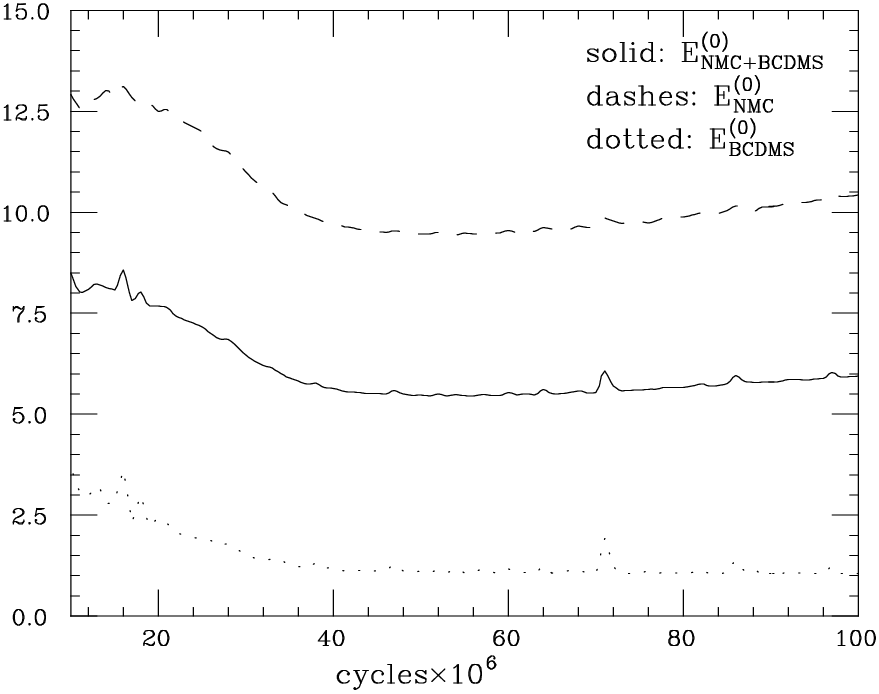}
\caption{Deuteron training with BCDMS data.}
\end{minipage}
\end{figure}
\end{center}

\begin{figure}
\begin{center}
\epsfig{width=0.6\textwidth,figure=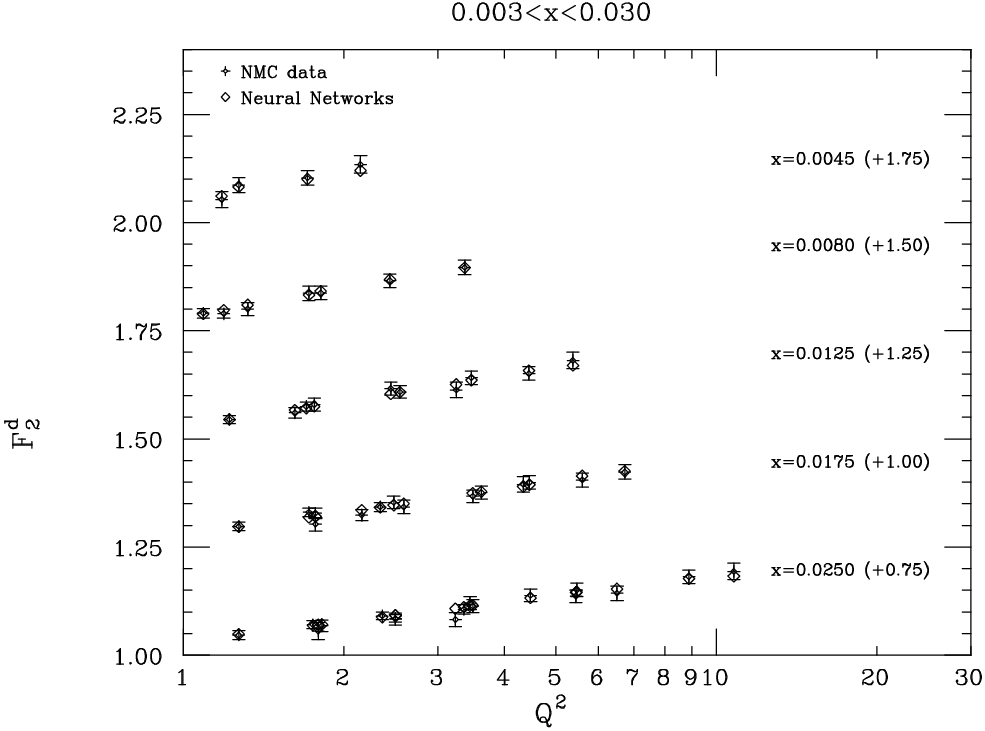} 
\epsfig{width=0.6\textwidth,figure=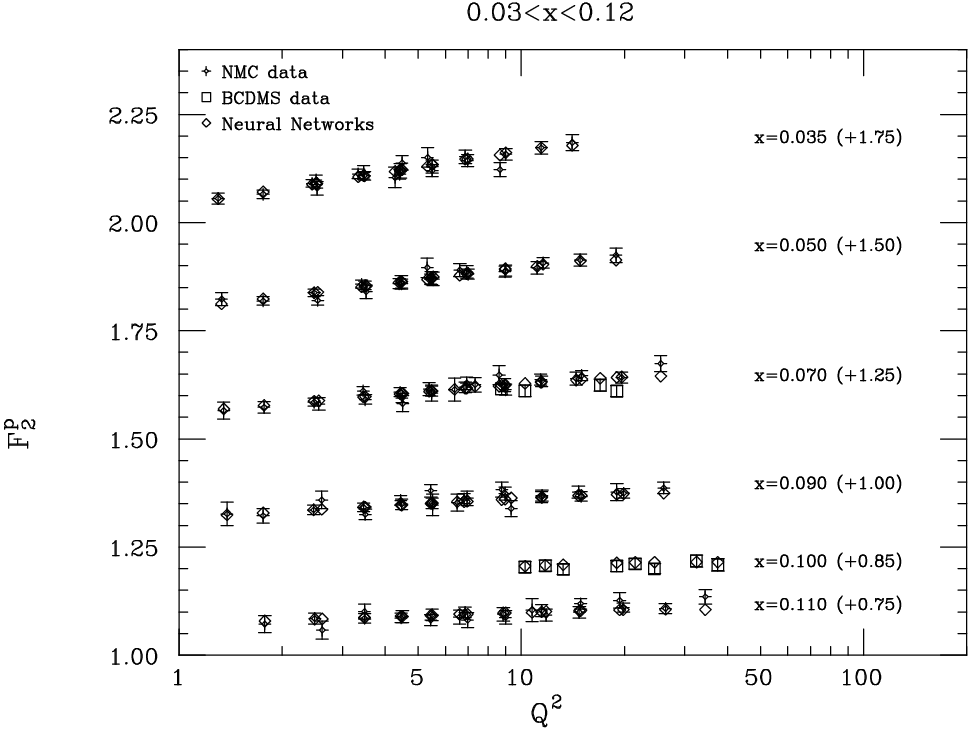} 
\epsfig{width=0.6\textwidth,figure=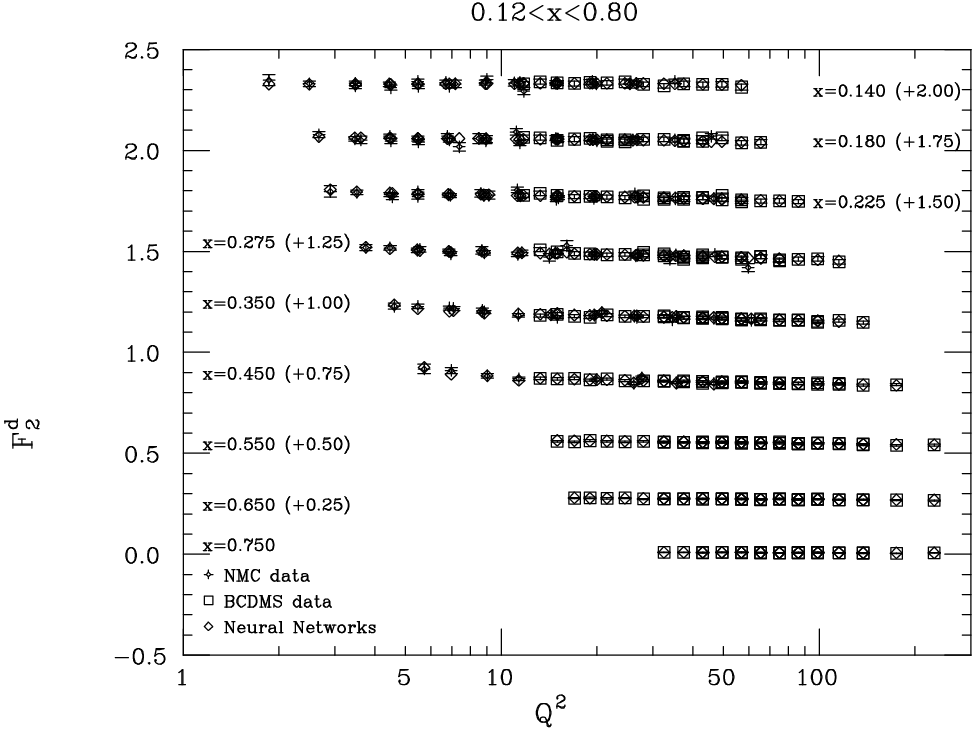} 
\end{center}
\caption{Deuteron data and network prediction at the corresponding $(x,Q^2)$
value.}  
\end{figure}

\begin{table}[t]  
\begin{center}  
\begin{tabular}{cccc} 
\multicolumn{4}{c}{}\\  
\hline
$N_{net}$ & NMC+BCDMS & NMC & BCDMS\\
\hline  
$\chi^{2}$ 
  & 1.10 & 1.16 & 1.03 \\ 
$\lan E \ran_{rep}$ 
  & 2.64 & 2.74 & 2.53 \\
${\cal R}$ 
  & 0.96 & 0.90 & 1.04\\
$r\l[F^{(net)}\r]$ 
  & 0.998 & 0.986 &  0.999\\
\hline
$\Big\lan V[\sigma^{(net)}]\Big\ran_{dat}$ 
  & $9.1\times 10^{-5}$ & $1.4\times 10^{-4}$ & $3.2\times 10^{-5}$  \\
$\Big\lan PE[\sigma^{(net)}]\Big\ran_{dat}$ 
  & 61\%  & 66\% & 56\%\\
$\Big\lan \sigma^{(exp)} \Big\ran_{dat}$ 
  & 0.010 & 0.014 & 0.006 \\
$\Big\lan \sigma^{(net)} \Big\ran_{dat}$ 
  & 0.006  & 0.007 & 0.004 \\
$r[\sigma^{(net)}]$ 
  & 0.72  & 0.34 &  0.94\\
\hline
$\Big\lan V[\rho^{(net)}]\Big\ran_{dat}$ 
  & 0.17 & 0.21 &  0.11\\
$\Big\lan \rho^{(exp)} \Big\ran_{dat}$ 
  & 0.31 & 0.22 & 0.43 \\
$\Big\lan \rho^{(net)} \Big\ran_{dat}$ 
  & 0.63 & 0.57 &  0.71\\
$r[\rho^{(net)}]$ 
  & 0.50 & 0.27 &  0.69\\
\hline
$\Big\lan V[\mathrm{cov}^{(net)}]\Big\ran_{dat}$ 
  & $3.4\times 10^{-10}$ & $ 1.2\times 10^{-9}$ & $ 2.1\times 10^{-10}$ \\
$\Big\lan \mathrm{cov}^{(exp)} \Big\ran_{dat}$ 
  & $3.2\times 10^{-5}$ & $4.0\times 10^{-5}$  & $2.2\times 10^{-5}$ \\
$\Big\lan \mathrm{cov}^{(net)} \Big\ran_{dat}$ 
  & $2.6\times 10^{-5}$ & $3.0\times 10^{-5}$  & $1.3\times 10^{-5}$ \\
$r[\mathrm{cov}^{(net)}]$ 
  & 0.71 & 0.63 & 0.90\\
\hline
\end{tabular}
\end{center}
\caption{Deuteron results}
\end{table}
The final training parameters are chosen 
as follows:   $9.6\times
10^6$ cycles with learning rate $\eta=9\times 10^{-3}$ and error function
Eq.~(\ref{simpenergy}), followed by $2\times 9\times10^7$ cycles
with $\eta=4\times
10^{-8}$  and error
function
Eq.~(\ref{diaenergy}). The momentum term Eq.~(\ref{momterm}) is always
set at $\alpha=0.9$. A sample of 1000 networks trained in this way is
used to produce values of the structure function and error which are
compared to the data in 
Figure~22. The features of these networks are
summarized in Table~6.

The results of Table~6 show that central values and errors
are well reproduced. There is a moderate amount of
variance reduction for BCDMS data, and a somewhat larger reduction for
NMC. The covariance is well reproduced, indicating that systematics
are reproduced in a faithful way. The features interpolation and
extrapolation are similar as in the nonsiglet case: interpolation and
extrapolation at large $x$ where small $x$ data are available are
quite accurate, extrapolation at large $Q^2$ by a small amount is
tolerable, while extrapolation at small $Q^2$ or at small $x$  are
subject to large uncertainties.
\vfill\eject
\section{Summary}

We have presented a determination of the probability density in the
space of structure functions for the structure function $F_2$ for
proton, deuteron and nonsinglet, as determined from experimental data
of the NMC and BCDMS collaborations. Our results, for each of the
three structure functions,  take the form of a
set of 1000 neural nets,
each of which gives a determination of $F_2$ for
given $x$ and $Q^2$. The distribution of these functions is a Monte
Carlo sampling of the probability density. 
This Monte Carlo sampling has been obtained by first, producing a 
sampling of  the space of
data points based on the available experimental information through a
set of Monte Carlo replicas of the original data, and then, 
training each neural net to one of these replicas.

In practice, all functions are given by a FORTRAN routine  which
reproduces a feed--forward neural network (described in Section~3)
entirely determined by a set of 47 real parameters.
Each function is then specified by the set of values for these
parameters.
Our results are available at the web page {\tt
http://sophia.ecm.ub.es/f2neural/}. The full set of FORTRAN routines
and parameters can be downloaded from this page. 
On--line plotting and computation facilities for  structure functions, errors
and point--to--point correlations are also available through this web
interface. 

Given the Monte Carlo sample of the probability density, any
functional of the structure function, as well as errors and
correlations,  can be computed by averaging over
the sample according to Eq.~(\ref{discfunave}). 
The sample has been carefully tested
against the available data to
produce reliable results wherever data are available. Because the data
provide a rather fine scanning of the $x,Q^2$ plane (Figure 1), this
representation of the probability density is expected to be very
accurate throughout the region of the data. However, care should be
adopted in extrapolating results far from the data region. On the
other hand, we have verified that as one gets further from the data
region, the spread of the probability density rapidly increases:
so, errors computed from the Monte Carlo sample should be a good
indicator of the point where the extrapolation becomes unreliable.

Our results can be useful for a variety of applications to the
precision phenomenology of deep--inelastic scattering. 
A particularly promising application consists of coupling the
bias--free determination of structure functions discussed here with
the bias--free technique for the description of QCD scaling violations
based on the use of truncated moments~\cite{trmom}. In this
approach,  QCD
evolution equations are directly expressed in terms of the scale
dependence of the contribution to moments of structure functions from
the experimentally accessible region. A precision
determination of the strong coupling $\alpha_s$ based on these
techniques will be presented in a companion paper~\cite{prep}.

In conclusion, this is the first complete determination of a
probability density in a space of structure functions. As such, it
provides an example of
a  bias--free determination of a function
with its associate error from a discrete set of experimental data,
which is a
problem of foremost importance in the determination of errors on
parton distributions.
\bigskip

{\bf Acknowledgements:} We thank F.~Zomer and A.~Milsztajn for patiently
explaining to us the structure of the data and their errors, and
G.~d'Agostini, R.~Ball, L.~Magnea  and G.~Ridolfi for several
discussions along the
course of this work. We are very grateful to L.~Magnea for a critical
reading of the manuscript, to R.~Ball for stimulating comments, and to
S.~Simula for trying out the web interface.
This work was supported in part by
EU TMR  contract FMRX-CT98-0194 (DG 12 - MIHT)
and the Spanish and Catalan grants AEN99-0766, AEN99-0483, 1999SGR-00097.


\end{document}